\documentclass[aps,prd,twocolumn,nofootinbib]{revtex4-1}

\usepackage{graphicx}
\usepackage{hyperref}
\usepackage{slashed}
\usepackage{color}

\newcommand{\slashp}[1]{\not{\!#1}}

\begin{document}

\title{Excited Heavy Quarkonium Production at the LHC through $W$-Boson Decays}
\author{Qi-Li Liao$^{1}$}
\author{Xing-Gang Wu$^{1,2}$}
\email{wuxg@cqu.edu.cn}
\author{Jun Jiang$^{1}$}
\author{Zhi Yang$^{1}$}
\author{Zhen-Yun Fang$^{1}$}
\author{Jia-Wei Zhang$^{3}$}

\address{$^{1}$ Department of Physics, Chongqing University, Chongqing 401331, P.R. China \\
$^{2}$ SLAC National Accelerator Laboratory, 2575 Sand Hill Road, Menlo Park, CA 94025, USA \\
$^3$ Department of Physics, Chongqing University of Science and Technology, Chongqing 401331, P.R. China}

\date{\today}

\begin{abstract}

Sizable amount of heavy-quarkonium events can be produced through $W$-boson decays at the LHC. Such channels will provide a suitable platform to study the heavy-quarkonium properties. The ``improved trace technology", which disposes the amplitude ${\cal M}$ at the amplitude-level, is helpful for deriving compact analytical results for complex processes. As an important new application, in addition to the production of the lower-level Fock states $|(Q\bar{Q'})[1S]\rangle$ and $|(Q\bar{Q'})[1P]\rangle$, we make a further study on the production of higher-excited $|(Q\bar{Q'})\rangle$-quarkonium Fock states $|(Q\bar{Q'})[2S]\rangle$, $|(Q\bar{Q'})[3S]\rangle$ and $|(Q\bar{Q'})[2P]\rangle$. Here $|(Q\bar{Q'})\rangle$ stands for the $|(c\bar{c})\rangle$-charmonium, $|(c\bar{b})\rangle$-quarkonium and $|(b\bar{b})\rangle$-bottomonium respectively. We show that sizable amount of events for those higher-excited states can also be produced at the LHC. Therefore, we need to take them into consideration for a sound estimation. If assuming all excited heavy-quarkonium states decay to the ground spin-singlet $[1{^1S_0}]$-wave state with $100\%$ efficiency via electromagnetic or hadronic interactions, we obtain the total decay width by adding all the mentioned Fock states together; i.e. by taking the bound-state parameters under the Buchm\"{u}ller-Tye potential model, we obtain $\Gamma_{W^+\to |(c\bar{c})\rangle +c\bar{s}} =591.3$ KeV, $\Gamma_{W^+\to |(c\bar{b})\rangle+b\bar{s}} =27.0$ KeV,  $\Gamma_{W^+\to |(c\bar{b})\rangle +c\bar{c}}= 2.01$ KeV, and $\Gamma_{W^+\to |(b\bar{b})\rangle +c\bar{b}} = 93.3 $ eV. At the LHC with the luminosity ${\cal L}\propto 10^{34}cm^{-2}s^{-1}$ and the center-of-mass energy $\sqrt{S}=14$ TeV, this shows that $8.7\times10^6$ $\eta_c$ and $J/\Psi$, $4.3\times10^5$ $B_c$ and $B^*_c$, $1.4\times10^3$ $\eta_b$ and $\Upsilon$ events per year can be obtained through $W^+$ decays. \\

\noindent {\bf PACS numbers:} 12.38.Bx, 14.40.Pq, 14.70.Fm

\end{abstract}

\maketitle

\section{Introduction}

Within the framework of Non-Relativistic QCD (NRQCD)~\cite{nrqcd}, a doubly heavy meson is considered as an expansion of Fock states. A systematic study on the production of lower $|(Q\bar{Q'})\rangle$-quarkonium Fock states at the $1S$-level ($|(Q\bar{Q'})[1^{1}S_{0}]\rangle$ and $|(Q\bar{Q'})[1^{3}S_{1}]\rangle$) and the $1P$-level ($|(Q\bar{Q'})[1^{1}P_{1}]\rangle$ and $|(Q\bar{Q'})[1^{3}P_{J}]\rangle (J=0, 1, 2)$) via the $W^+$ semi-inclusive decays has been done in Ref.\cite{wsp}. Here $|(Q\bar{Q'})\rangle$-quarkonium stands for the $|(c\bar{c})\rangle$-charmonium, $|(c\bar{b})\rangle$-quarkonium and $|(b\bar{b})\rangle$-bottomonium respectively. At the LHC, due to its high collision energy and high luminosity, sizable amount of heavy-quarkonium events can be produced through $W^+$ decays \cite{wsp,w}. So these channels maybe an important supplement for other measurements at the LHC.

In this paper, we will make a further discussion on the production of even higher $|(Q\bar{Q'})\rangle$-quarkonium Fock states via the $W^+$ semi-inclusive decays, which are at the $nS$-level ($|(Q\bar{Q'})[nS]\rangle$) and $nP$-level ($|(Q\bar{Q'})[nP]\rangle$) with $n\geq2$ respectively. As shown in Ref.\cite{wsp}, due to the color-suppression of the amplitude and the relative velocity suppression of the color-octet matrix element, the color-octet $(Q\bar{Q'})$-component provides negligible contributions, so at the present, we will only discuss the color-singlet states' production channels. Moreover, because the $W$ decay-widths to the $|(b\bar{b})\rangle$-bottomonium states are quite small, we will concentrate our attention on the $|(c\bar{c})\rangle$-charmonium and the $|(c\bar{b})\rangle$-quarkonium production.

For the purpose, we will deal with the $1\to 3$ decay channel, $W^{+} \to |(Q\bar{Q'})[n]\rangle +q+ \bar{q}'$, which is a short notation for the following four semi-inclusive decay channels: $ W^{+}\rightarrow |(c \bar{c})[n]\rangle+c \bar{s}$, $W^{+}\rightarrow |(c\bar{b})[n]\rangle+b \bar{s}$, $W^{+}\rightarrow |(c\bar{b})[n]\rangle+c\bar{c}$ and $ W^{+}\rightarrow |(b \bar{b})[n]\rangle+c \bar{b}$, respectively. Here $n$ stands for the corresponding quantum number of the heavy $|(Q\bar{Q'})\rangle$-quarkonium Fock state, $q$ and $\bar{q}'$ stand for the outgoing quark and anti-quark. Intuitively, the heavy-quarkonium production could be understood in terms of two distinct steps: one is the production of the heavy $(Q\bar{Q'})$-pair with certain quantum number $n$, together with the out-going quark $q$ and anti-quark $\bar{q}'$; and the second subsequent step is the evolution of such $Q\bar{Q'}$ pair into a quarkonium bound state. The first step is pQCD calculable, where the quantum state $[n]$ of the $(Q\bar{Q'})$-pair can be achieved by a suitable projector~\cite{projector1,projector2,projector3}, and the second step is characterized by a non-perturbative matrix element which is proportional to the inclusive transition probability of the perturbative $(Q\bar{Q'})$-pair with certain quantum number $[n]$ into a bound state $|(Q\bar{Q'})[n]\rangle$~\cite{nrqcd}.

To deal with the heavy-quarkonium production through the $W^+$ semi-inclusive decays, one needs to derive the pQCD calculable squared amplitude. The usual way is to deal with the squared amplitude $|{\cal M}|^2$. Because of the emergence of massive-fermion lines, the analytical expression for the squared amplitude becomes too complex and lengthy for more (massive) particles in the final states and for higher-level Fock states to be generated. For example, to derive the amplitudes for the $P$-wave states, one also needs to get the derivative of the amplitudes over the relative momentum of the constitute quarks. It has been found that to do the numerical calculation using this conventional squared-amplitude technology becomes time-consuming for those complex processes, since the cross-terms of the matrix elements increase with the increment of Feynman diagrams, $|{\cal M}|^2=\sum_{ij}{\cal M}_{i} {\cal M}^*_{j}$, where $i$ and $j$ stand for the number of Feynman diagrams of the process.

One important way to solve this is to deal with the process directly at the amplitude level. For the purpose, the ``improved trace technology" is suggested and developed in the literature \cite{itt0,itt1,itt2,itt3,itt4}. After generating proper phase-space points, one first calculate the numerical value for the amplitudes ${\cal M}_{i}$, and then sum these values algebraically and square it to get the squared amplitude, $|{\cal M}|^2=|\sum_{i}{\cal M}_{i}|^2$; through such way, numerical simulation efficiency can be greatly improved in comparison to the conventional squared-amplitude technology. Moreover, under the approach, many simplifications can be done at the amplitude level due to the fermion-line symmetries and the specific properties of each heavy-quarkonium Fock states, then, we can even written down the analytic expressions for the amplitude.

The remaining parts of this paper are organized as follows: In Sec.II, we give a short review of dealing the $W^+$ semi-inclusive decays by using the ``improved trace technology". In Sec.III, we present the numerical results and discuss on the properties of the heavy-quarkonium production through $W^+$ decays. The final section is reserved for a summary.

\section{a short review of the calculation technology}

Our present method for deriving the decay width of $W^{+}(k) \to |(Q\bar{Q'})[n]\rangle(q_3) +q(q_2)+ \bar{q}'(q_1)$, especially the pQCD calculable part, is the same as that of Ref.\cite{wsp}, for self-consistency, we will list the main points here.

According to the NRQCD factorization formula \cite{petrelli}, the differential decay width of $W^{+}(k) \to |(Q\bar{Q'})[n]\rangle(q_3) +q(q_2)+ \bar{q}'(q_1)$ can be factorized as
\begin{equation}
d\Gamma=\sum_{n} d\hat\Gamma(W^+ \to (Q\bar{Q'})[n]+q \bar{q}') \langle{\cal O}^H(n) \rangle ,
\end{equation}
where $\langle{\cal O}^{H}(n)\rangle$ describes the hadronization of a $Q\bar{Q'}$-pair into the observable quark state $H$ and is proportional to the transition probability of the perturbative state $(Q\bar{Q'})[n]$ into the bound state $|(Q\bar{Q'})[n]\rangle$. The parameters $k$ and ${q_i}$ are momenta of the corresponding out-going particles.

The short-distance differential decay width
\begin{equation}
d\hat\Gamma(W^{+}\to (Q\bar{Q}')[n]+q\bar{q}')= \frac{1}{2k^0} \overline{\sum} |{\cal M}|^{2} d\Phi_3,
\end{equation}
where $\overline{\sum}$ means that we need to average over the spin states of the initial particles and to sum over the color and spin of all the final particles. In the $W^+$ rest frame, the three-particle phase space can be written as
\begin{equation}
d{\Phi_3}=(2\pi)^4 \delta^{4}\left(k - \sum_f^3 q_{f}\right)\prod_{f=1}^3 \frac{d^3{\vec{q}_f}}{(2\pi)^3 2q_f^0}.
\end{equation}
The $1 \to 3$ phase space with massive quark/antiqark in the final state can be found in Refs.\cite{itt2,itt3}. With the help of the formulas listed in Refs.\cite{itt2,itt3}, one can not only derive the whole decay width but also obtain the corresponding differential decay widths that are helpful for experimental studies, such as $d\Gamma/ds_1$, $d\Gamma/ds_2$, $d\Gamma/d\cos\theta_{13}$ and $d\Gamma/d\cos\theta_{23}$, where $s_1=(q_1+q_3)^2$, $s_2=(q_1+q_2)^2$, $\theta_{13}$ is the angle between $\vec{q}_1$ and $\vec{q}_3$ in the $W^+$ rest frame, and $\theta_{23}$ is the angle between $\vec{q}_2$ and $\vec{q}_3$ in the $W^+$ rest frame. Especially, the partial decay-widths over $s_1$ and $s_2$ can be expressed as:
\begin{eqnarray}
\frac{d\Gamma}{ds_1 ds_2} &=& \frac{\langle{\cal O}^H_1(n) \rangle}{256 \pi^3m_W^3} \overline{\sum}|{\cal M} |^2 ,
\end{eqnarray}
where $m_W$ stands for the $W$-boson mass. The integration over $s_1$ and $s_2$ can be done with the help of the VEGAS program \cite{vegas} \footnote{The improved VEGAS version can be found in the programs BCVEGPY \cite{bcvegpy} and GENXICC \cite{genxicc}.}.

Selection of appropriate angular momentum quantum number for the quarkonium state is done by suitable projector and by performing proper polarization sum \cite{projector1,projector2,projector3}. The color-singlet non-perturbative matrix-element $\langle{\cal O}^H_1 (n) \rangle$ can be related to the Schr\"{o}dinger wavefunctions at the origin $|\psi_{(Q\bar{Q'})}(0)|$ or the first derivative of the wavefunctions at the origin $|\psi^{'}_{(Q\bar{Q'})}(0)|$. For the color-singlet $[nS]$- and $[nP]$- wave states, we have \cite{projector2,geg}
\begin{eqnarray}
\langle{\cal O}^H_1 (nS) \rangle &\simeq& |\psi_{|(Q\bar{Q'})[nS]\rangle}(0)|^2 , \\
\langle{\cal O}^H_1 (nP) \rangle &\simeq& |\psi^{'}_{|(Q\bar{Q'})[nP]\rangle}(0)|^2.
\end{eqnarray}
Here we have adopted the convention of Ref.\cite{projector2} for the non-perturbative matrix elements. Since the spin-splitting effects are small, we will not distinguish the difference between the wavefunction parameters for the spin-singlet and spin-triplet states at the same $n$-{\it th} level.

\begin{figure}[ht]
\includegraphics[width=0.45\textwidth]{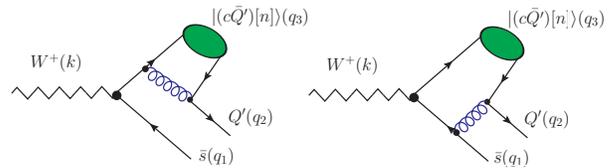}
\caption{Feynman diagrams for the process $W^+(k)\rightarrow |(c\bar{Q'})[n]\rangle(q_3) + Q'(q_2) \bar{s}(q_1)$, where $Q'$ stands for $c$ or $b$ quark accordingly, and $|(c\bar{Q'})[n]\rangle$ stands for a heavy-quarkonium Fock state. } \label{feyn1}
\end{figure}

\begin{figure}[ht]
\includegraphics[width=0.45\textwidth]{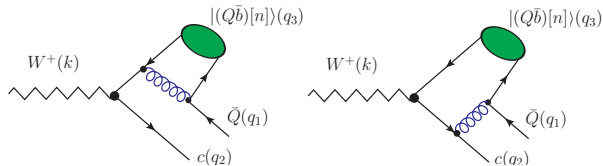}
\caption{Feynman diagrams for the process $W^+(k)\rightarrow |(Q\bar{b})[n]\rangle(q_3) + c(q_2) \bar{Q}(q_1)$, where $Q$ stands for the $c$ or $b$ quark accordingly, and $|(Q\bar{b})[n]\rangle$ stands for a heavy-quarkonium Fock state. } \label{feyn2}
\end{figure}

The amplitude ${\cal M}$ for the mentioned channels can be calculated from two types of Feynman diagrams which are shown in Figs.(\ref{feyn1},\ref{feyn2}) respectively. We present the two process $W^{+}\rightarrow |(c\bar{c})[n]\rangle + c \bar{s}$ and $W^{+}\rightarrow |(c\bar{b})[n]\rangle + b \bar{s}$ as $W^+(k) \rightarrow |(c\bar{Q'})[n]\rangle(q_3) + Q'(q_2)\bar{s}(q_1)$, where $Q'$ stands for $c$ or $b$ quark accordingly. We present the two processes $W^{+}\rightarrow |(c\bar{b})[n]\rangle +c\bar{c}$ and $W^{+}\rightarrow |(b\bar{b})[n]\rangle +c \bar{b}$ as $W^+(k) \rightarrow |(Q\bar{b})[n]\rangle(q_3) + c(q_2)\bar{Q}(q_1)$, where $Q$ stands for the $c$ or $b$ quark accordingly. In those Feynman diagrams, the intermediate gluon should be hard enough to produce a $c\bar{c}$ pair or $b\bar{b}$ pair, so the amplitude is perturbative QCD calculable.

For each production channel, the pQCD calculable amplitude for a specific spin-state combination ${\cal M}_{s s'}$ can be generally expressed as
\begin{equation}
i{\cal M}_{s s'} = {\cal{C}} {\bar u_{s i}}({q_2}) \sum\limits_{n = 1}^{2} {{\cal A} _n } {v_{s' j}}({q_1}),
\end{equation}
where $s$ and $s'$ are spin indices, $i$ and $j$ are color indices for the outgoing quark and antiquark. The overall factor ${\cal C}=\frac{2gg_s^2 V_{CKM}}{3\sqrt{6}}\delta_{ij}$. $V_{CKM}$ stands for the Cabibbo-Kobayashi-Maskawa matrix element, $V_{CKM}=V_{cs}$ for $W^{+}\rightarrow |(c\bar{c})[n]\rangle + c \bar{s}$ and $W^{+}\rightarrow |(c\bar{b})[n]\rangle + b \bar{s}$; $V_{CKM}=V_{cb}$ for $W^{+}\rightarrow |(c\bar{b})[n]\rangle +c\bar{c}$ and $W^{+}\rightarrow |(b\bar{b})[n]\rangle +c \bar{b}$.

The expressions for ${{\cal A} _n }$ can be read from Figs.(\ref{feyn1},\ref{feyn2}), and then we deal with the amplitude ${\cal M}$ by using the ``improved trace technology" \cite{itt0,itt1,itt2,itt3,itt4}. Under such approach, we first arrange the amplitude into four orthogonal sub-amplitudes ${\cal M}_{\pm{s}\pm{s'}}$ according to the spins of the outgoing quark $q$ with spin $s$ and antiquark $\bar{q}'$ with spin $s'$. After summing up the spin states of the outgoing quark/antiquark, the squared amplitude can be divided into four orthogonal parts,
\begin{equation}
|{\cal M}|^2 = |{\cal M}_{1}|^2 + |{\cal M}_{2}|^2 + |{\cal M}_{3}|^2 + |{\cal M}_{4}|^2,
\end{equation}
where the four amplitudes ${\cal M}_i$ $(i=1,\cdots,4)$ are defined as
\begin{eqnarray}
{\cal M}_1 &=& \frac{{\cal M}_{ss'} +{\cal M}_{-s-s'}}{\sqrt{2}} \;,\;
{\cal M}_2 = \frac{{\cal M}_{ss'} -{\cal M}_{-s-s'}}{\sqrt{2}} \;,\nonumber\\
{\cal M}_3 &=& \frac{{\cal M}_{s-s'}-{\cal M}_{-ss'}}{\sqrt{2}} \;,\;
{\cal M}_4 = \frac{{\cal M}_{s-s'} +{\cal M}_{-ss'}}{\sqrt{2}} \;.
\end{eqnarray}
These four amplitudes ${\cal M}_i$ can be transformed into trace forms by properly dealing with the massive spinors with the help of an arbitrary light-like momentum $k_0$ and an arbitrary space-like momentum $k_1$ ($k_1^2=-1$), $k_0$ and $k_1$ satisfy the constraint $k_0\cdot k_1 =0$.

More explicitly, we first introduce a massless spinor with negative helicity $u_{-}(k_0)$ which satisfies the following projection
\begin{equation}
u_{-}(k_0)\bar{u}_{-}(k_0)=\omega_{-} \slashp{k_0} ,
\end{equation}
where $\omega_{-}=(1-\gamma_5)/2$. Then, we construct the positive helicity state as
\begin{equation}
u_{+}(k_0)=\slashp{k_1} u_{-}(k_0) .
\end{equation}
It is easy to check that $u_{+}(k_0)\bar{u}_{+}(k_0)=\omega_{+}\slashp{k_0}$, where $\omega_{+}=(1+\gamma_5)/2$. Using these two massless spinors, the massive spinors for the fermion and antifermion can be written as follows:
\begin{eqnarray}
u_s(q)&=&(\slashp{q}+m)u_{-}(k_0)/\sqrt{2k_0\cdot q} , \\
u_{-s}(q)&=&(\slashp{q}+m)u_{+}(k_0)/\sqrt{2k_0\cdot q} , \\
v_{s}(q)&=&(\slashp{q}-m)u_{-}(k_0)/\sqrt{2k_0\cdot q} , \\
v_{-s}(q)&=&(\slashp{q}-m)u_{+}(k_0)/\sqrt{2k_0\cdot q} ,
\end{eqnarray}
where the spin vector $s_\mu=\frac{q_\mu}{m}-\frac{m}{q\cdot k_0}k_{0\mu}$.

Using the above relations, the amplitudes ${\cal M}_i$ can be conveniently transformed into trace forms. And then we do the trace of the Dirac $\gamma$-matrix strings which will result in explicit series over some limited and independent Lorentz-structures. The final results are independent of the choice of $k_0$ and $k_1$, one can choose them to be those that can maximumly simply the analytical expressions of the amplitude. The compact and complex expressions for the amplitudes of the $S$-wave and $P$-wave cases can be found in the appendix of Ref.\cite{wsp}.

These results provide the foundation for estimating the higher Fock-states' properties.

Even though it is hard to get the analytic results for the conventional squared-amplitude technology, as a cross-check, we also adopt it to calculate the $W^+$-decay widths numerically. Under the same parameter values, we have found agreement of these two approaches.

\section{numerical results and discussions}

\subsection{Input parameters}

When doing the numerical calculation, the input parameters are chosen as the following values \cite{wtd,pdg}: $m_W=80.399$ GeV, $\Gamma_{W^+}=2.085$ GeV, $m_s=0.105$ GeV, $|V_{cs}|=1.023\pm0.036$ and $|V_{cb}|=0.0406\pm0.0013$. Leading-order $\alpha_s$ running is adopted. An optimal process to set the renormalization scale $\mu_r$ has recently been suggested in the literature, i.e. the Principle of Maximum Conformality (PMC) \cite{pmc}, which is renormalization-scheme independent and can provide almost renormalization-scale independent estimation even at the fixed-order. However to set the leading-order PMC scale, one needs to know the next-to-leading order $\{\beta_i\}$-terms which will be absorbed into the $\alpha_s$-running to form a commensurate leading-order PMC scale. For the present leading-order calculation, since $\Gamma\propto\alpha_s^2(\mu_r)$, the renormalization scale-uncertainty can be easily figured out when we know the physical scale well. For clarity, we set $\mu_r$ to the conventional choice of $2 m_c$, and $\alpha_s(2m_c)=0.26$.

\begin{table}
\begin{tabular}{|c||c|c|c|c|}
\hline
~~~ ~~~& ~~ $n=1$ ~~ &~~ $n=2$~~  &~~ $n=3$~~  \\
\hline
$|R_{|(c\bar{c})[nS]\rangle}(0)|^2$ & 0.810 ${\rm GeV}^3$ & 0.529 ${\rm GeV}^3$ & 0.455 ${\rm GeV}^3$ \\
$|R^{'}_{|(c\bar{c})[nP]\rangle}(0)|^2$ & 0.075 ${\rm GeV}^5$ & 0.102 ${\rm GeV}^5$ & ~$\sim$~ \\
\hline
\hline
$|R_{|(c\bar{b})[nS]\rangle}(0)|^2$ & 1.642 ${\rm GeV}^3$ & 0.983 ${\rm GeV}^3$ & 0.817  ${\rm GeV}^3$ \\
$|R^{'}_{|(c\bar{b})[nP]\rangle}(0)|^2 $ & 0.201 ${\rm GeV}^5$ & 0.264 ${\rm GeV}^5$ & ~$\sim$~ \\
\hline
$|R_{|(b\bar{b})[nS]\rangle}(0)|^2$ &6.477 ${\rm GeV}^3$ &3.234 ${\rm GeV}^3$ &2.474 ${\rm GeV}^3$ \\
$|R^{'}_{|(b\bar{b})[nP]\rangle}(0)|^2 $ &1.417 ${\rm GeV}^5$ & 1.653 ${\rm GeV}^5$ & ~$\sim$~ \\
\hline \hline
$m_c$& 1.48 GeV & 1.70 GeV & 1.90 GeV \\
\hline
$m_b$& 4.88 GeV & 5.00 GeV & 5.10 GeV \\
\hline
\end{tabular}
\caption{Bound-state parameters adopted in the calculation, which are derived under the Buchm\"{u}ller-Tye potential \cite{quigg}. }
\label{tabrpa}
\end{table}

As for the wavefunction at the origin and the first derivative of the wavefunction at the origin, we adopt the values derived in Ref.\cite{quigg} under the Buchm\"{u}ller-Tye-potential (BT-potential) model as their central values, since it is noted that the BT-potential has the correct two-loop short-distance behavior in pQCD \cite{wb}. The results for three other potential models, i.e. the Power-Law model \cite{am}, the Logarithmic model \cite{cj} and the Cornell model \cite{ektk}, will be adopted as an error analysis. Furthermore, similar to our previous treatment \cite{wsp}, we adopt the same constitute quark masses for the same $n$-{\it th} level Fock states. To ensure the gauge invariance of the hard amplitude, we set the $|(Q\bar{Q'})\rangle$-quarkonium mass $M$ to be $m_Q + m_{Q'}$.

We present the quarkonium bound-state parameters under the BT-potential model in Table~\ref{tabrpa}, where the constitute charm- and bottom- quark masses, the radial wavefunction at the origin and the first derivative of the radial wavefunctions at the origin for $|(Q\bar{Q'})[n]\rangle$-quarkonium are presented. Here $R_{|(Q\bar{Q'})[nS]\rangle}(0)$ and $R^{'}_{|(Q\bar{Q'})[nP]\rangle}(0)$ are related to the wavefunction at the origin and the first derivative of the wavefunction at the origin through the following equations:
\begin{eqnarray}
|\Psi_{|(Q\bar{Q'})[nS]\rangle}(0)| &=& \sqrt{{1}/{4\pi}} |R_{|(Q\bar{Q'})[nS]\rangle}(0)|, \nonumber\\
|\Psi^{'}_{|(Q\bar{Q'})[nP]\rangle}(0)| &=& \sqrt{{3}/{4\pi}} |R'_{|(Q\bar{Q'})[nP]\rangle}(0)| .
\end{eqnarray}

\subsection{Heavy-quarkonium production via $W^+$ decays}

\begin{table}
\begin{tabular}{|c||c|c|c|}
\hline
~~~~ & ~~n=1~~ & ~~n=2~~ & ~~n=3~~ \\
\hline
$W^+\rightarrow |(c\bar{c})[n^{1}S_{0}]\rangle+ c\bar{s}$ &132.0&56.36&22.85\\
\hline
$W^+\rightarrow |(c\bar{c})[n^{3}S_{1}]\rangle+ c\bar{s}$ &136.4&58.30&33.38\\
\hline\hline
$W^+\rightarrow |(c\bar{b})[n^{1}S_{0}]\rangle+b\bar{s}$ &6.39&3.53&2.74\\
\hline
$W^+\rightarrow |(c\bar{b})[n^{3}S_{1}]\rangle+b\bar{s}$ &5.49&3.07&2.41\\
\hline\hline
$W^+\rightarrow |(c\bar{b})[n^{1}S_{0}]\rangle+c\bar{c}$ &0.411&0.160&0.094\\
\hline
$W^+\rightarrow |(c\bar{b})[n^{3}S_{1}]\rangle+c\bar{c}$ &0.593&0.224&0.128\\
\hline
\end{tabular}
\caption{Decay widths (in unit: KeV) for the production of $|(Q\bar{Q'})[nS]\rangle$-quarkonium through $W^+$ decays.}
\label{tabrpb}
\end{table}

\begin{table}
\begin{tabular}{|c||c|c|}
\hline
~~~~ & ~~~n=1~~~ & ~~~n=2~~~ \\
\hline
$W^+\rightarrow |(c\bar{c}) [n^{1}P_{1}]\rangle+c\bar{s}$&22.95&12.98\\
\hline
$W^+\rightarrow |(c\bar{c}) [n^{3}P_{0}]\rangle+c\bar{s}$ &28.33&17.21\\
\hline
$W^+\rightarrow |(c\bar{c}) [n^{3}P_{1}]\rangle+c\bar{s}$ &28.31&19.07\\
\hline
$W^+\rightarrow |(c\bar{c}) [n^{3}P_{2}]\rangle+c\bar{s}$ &14.76&8.414\\
\hline\hline
$W^+\rightarrow |(c\bar{b}) [n^{1}P_{1}]\rangle+b\bar{s}$ &0.270&0.360\\
\hline
$W^+\rightarrow |(c\bar{b}) [n^{3}P_{0}]\rangle+b\bar{s}$ &0.733&0.693\\
\hline
$W^+\rightarrow |(c\bar{b}) [n^{3}P_{1}]\rangle+b\bar{s}$ &0.514&0.724\\
\hline
$W^+\rightarrow |(c\bar{b}) [n^{3}P_{2}]\rangle+b\bar{s}$ &0.034&0.039\\
\hline\hline
$W^+\rightarrow |(c\bar{b}) [n^{1}P_{1}]\rangle+c\bar{c}$ &0.105&0.066\\
\hline
$W^+\rightarrow |(c\bar{b}) [n^{3}P_{0}]\rangle+c\bar{c}$ &0.026&0.019\\
\hline
$W^+\rightarrow |(c\bar{b}) [n^{3}P_{1}]\rangle+c\bar{c}$ &0.054&0.036\\
\hline
$W^+\rightarrow |(c\bar{b}) [n^{3}P_{2}]\rangle+c\bar{c}$ &0.060&0.037\\
\hline
\end{tabular}
\caption{Decay widths (in unit: KeV) for the production of $|(Q\bar{Q'})[nP]\rangle$-quarkonium through $W^+$ decays.}
\label{tabrpc}
\end{table}

The decay widths for the aforementioned quarkonium states through the production channel, $W^{+} \to |(Q\bar{Q'})[n]\rangle +q  \bar{q'}$, are listed in Tables \ref{tabrpb} and \ref{tabrpc}.

From Tables \ref{tabrpb} and \ref{tabrpc}, it is found that in addition to the ground $1S$-level states, the higher $(Q\bar{Q'})$-quarkonium states can also provide sizable contributions to the total decay width; i.e.,
\begin{itemize}
\item For the charmonium production channel, $W^{+}\rightarrow |(c\bar{c})[n]\rangle + c \bar{s}$, the decay widths for $[n]=2S$, $3S$, $1P$ and $2P$-wave states are about $43\%$, $21\%$, $35\%$ and $21\%$ of that of the $1S$-level wave state, respectively.
\item For the $(c\bar{b})$-quarkonium production channel, $W^{+}\rightarrow |(c\bar{b})[n]\rangle + b\bar{s}$, the decay widths for $[n]=2S$, $3S$, $1P$ and $2P$-wave states are about $55\%$, $43\%$, $13\%$ and $15\%$ of that of the $1S$-level, respectively.
\item For the $(c\bar{b})$-quarkonium production channel, $W^{+}\rightarrow |(c\bar{b})[n]\rangle + c\bar{c}$, the decay widths for $[n]=2S$, $3S$, $1P$ and $2P$-wave states are about $38\%$, $22\%$, $24\%$ and $16\%$ of that of the $1S$-level, respectively.
\end{itemize}
Here, for convenience, we have used $[nS]$ to represent the summed decay width of $[n {^1S_0}]$ and $[n {^3S_1}]$ at the same $n$-{\it th} level; $[nP]$ to represent the summed decay width of $[n {^1P_1}]$ and $[n {^3P_J}]$ at the same $n$-{\it th} level.

\begin{figure}
\includegraphics[width=0.23\textwidth]{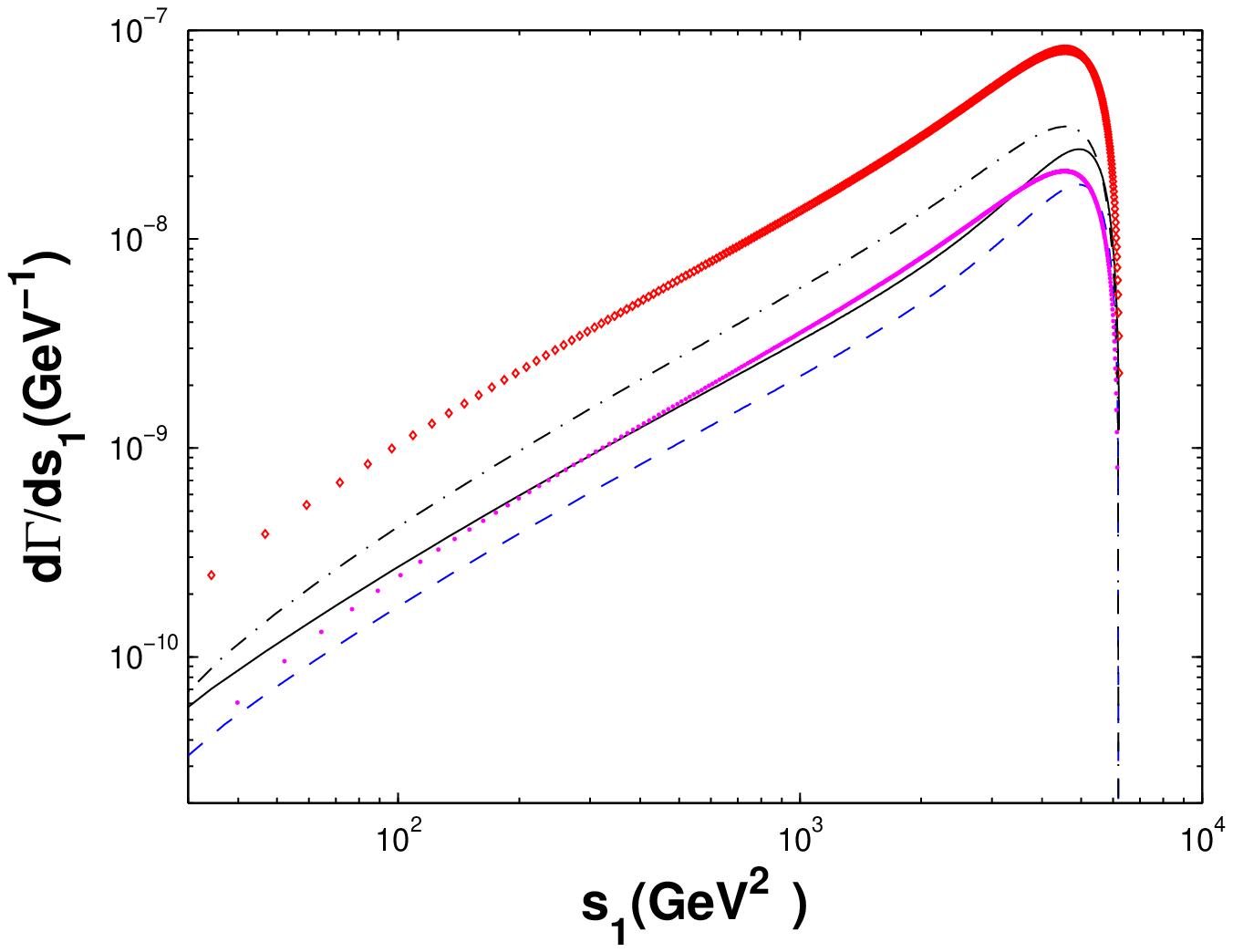}
\includegraphics[width=0.23\textwidth]{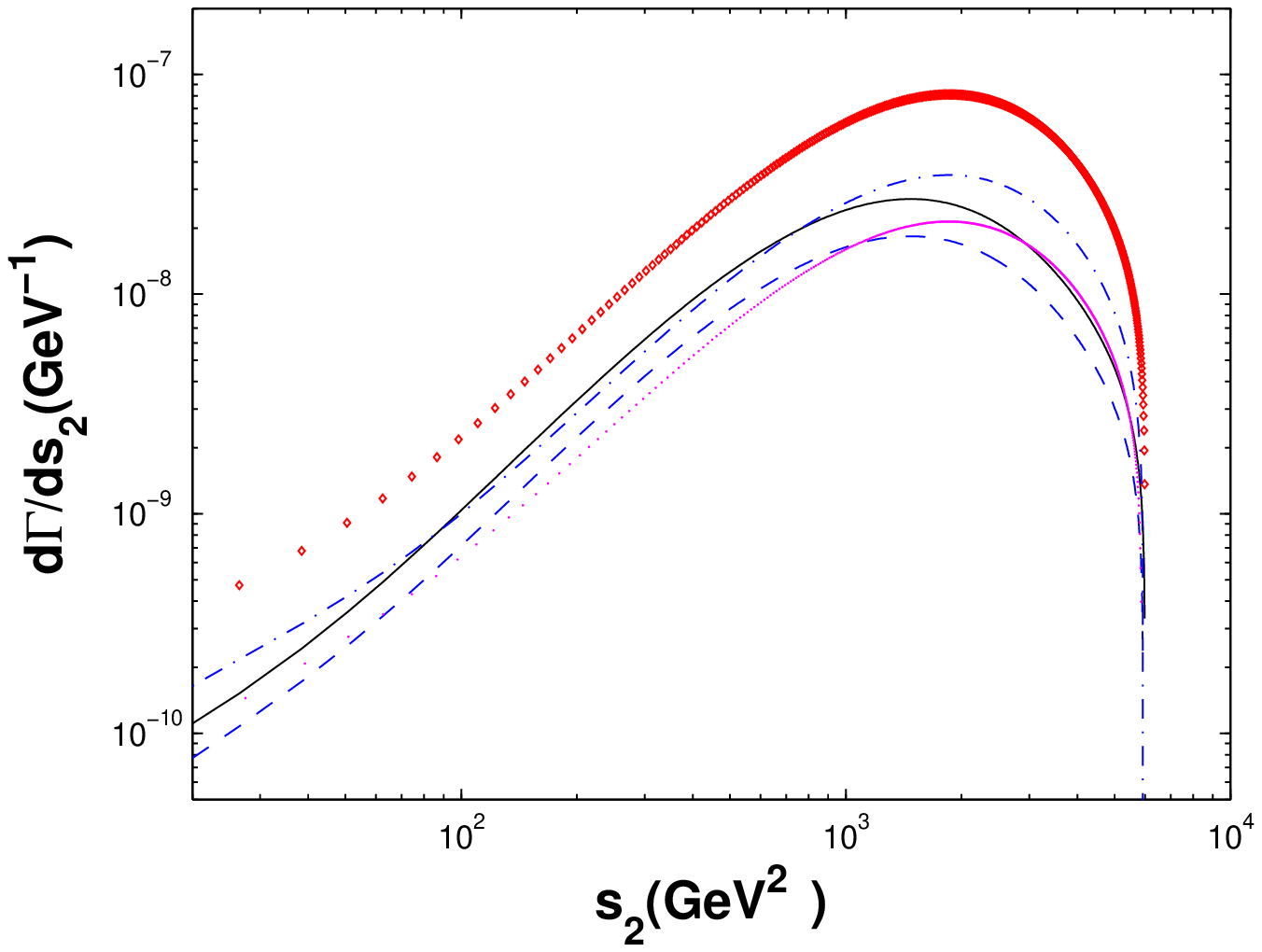}
\caption{Differential decay widths $d\Gamma/ds_1$ and $d\Gamma/ds_2$ for $W^+\rightarrow |(c\bar{c})[n]\rangle +c\bar{s}$, where the diamond, the dash-dot, the dotted, the solid and the dashed lines are for $|(c\bar{c})[1S]\rangle$, $|(c\bar{c})[2S]\rangle$, $|(c\bar{c})[3S]\rangle$, $|(c\bar{c})[1P]\rangle$ and $|(c\bar{c})[2P]\rangle$, respectively.} \label{CCcsdiss1s2}
\end{figure}

\begin{figure}
\includegraphics[width=0.23\textwidth]{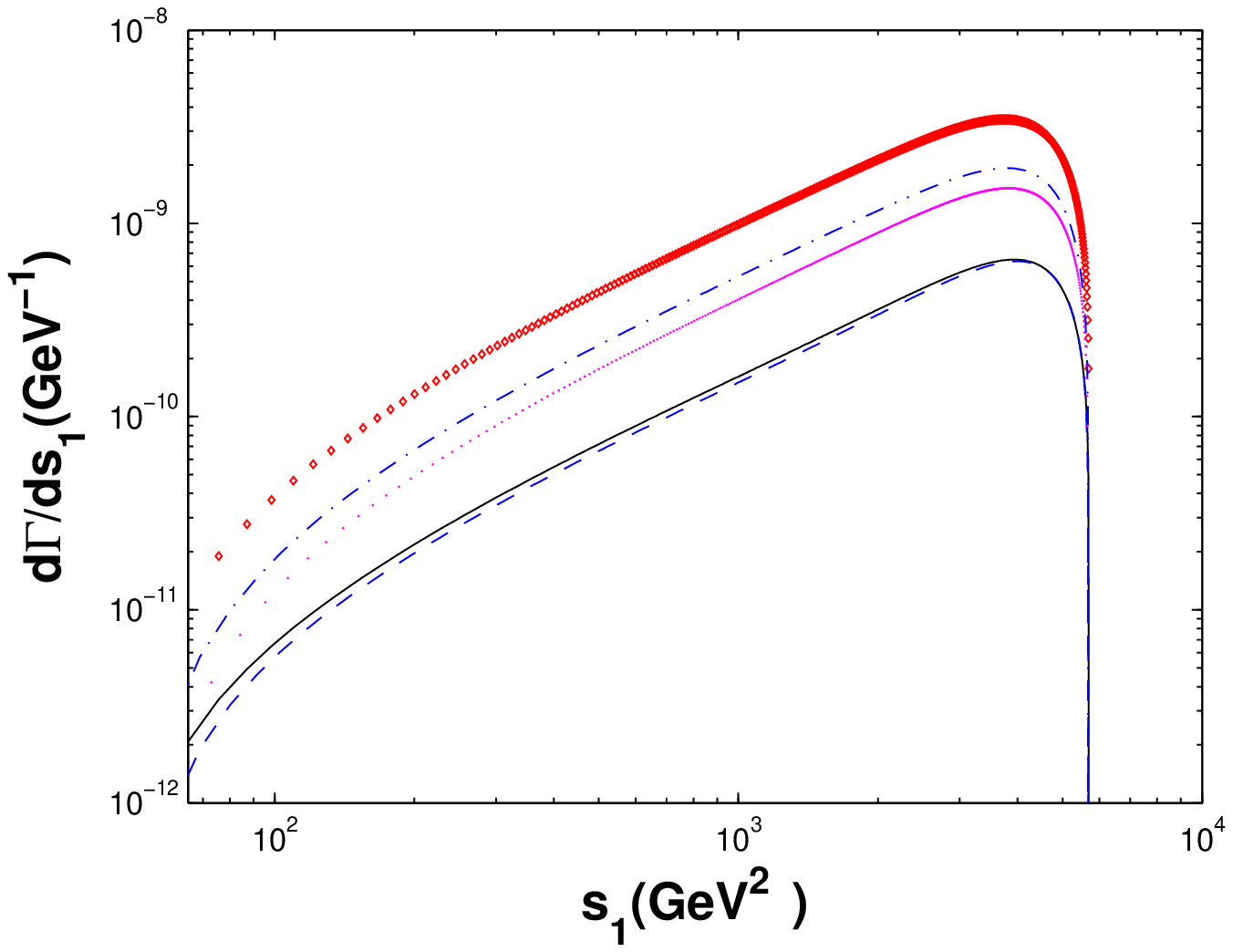}
\includegraphics[width=0.23\textwidth]{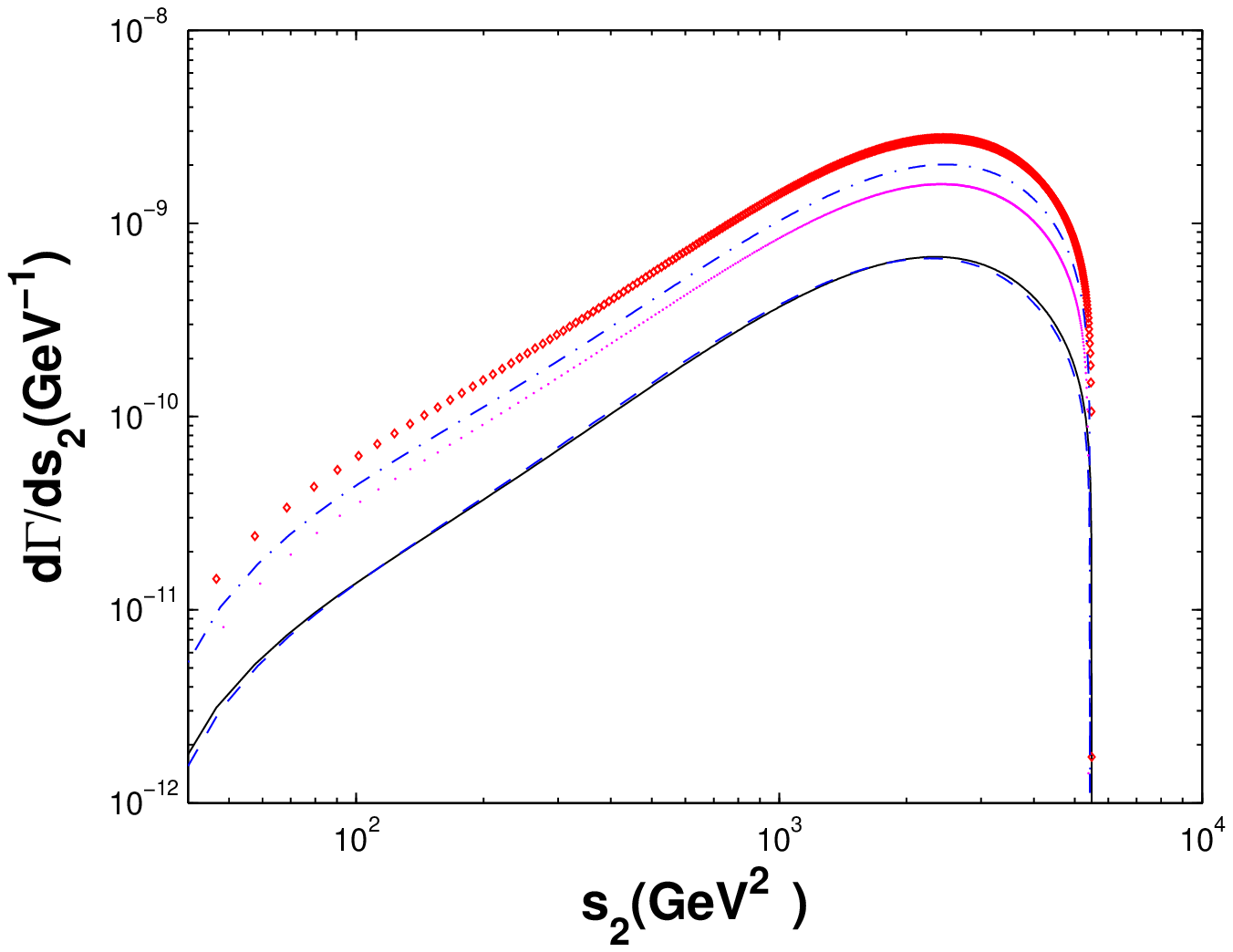}
\caption{Differential decay widths $d\Gamma/ds_1$ and $d\Gamma/ds_2$ for $W^+\rightarrow |(c\bar{b})[n]\rangle +b\bar{s}$, where the diamond, the dash-dot, the dotted, the solid and the dashed lines are for $|(c\bar{b})[1S]\rangle$, $|(c\bar{b})[2S]\rangle$, $|(c\bar{b})[3S]\rangle$, $|(c\bar{b})[1P]\rangle$ and $|(c\bar{b})[2P]\rangle$, respectively.} \label{Bcbsdiss1s2}
\end{figure}

\begin{figure}
\includegraphics[width=0.23\textwidth]{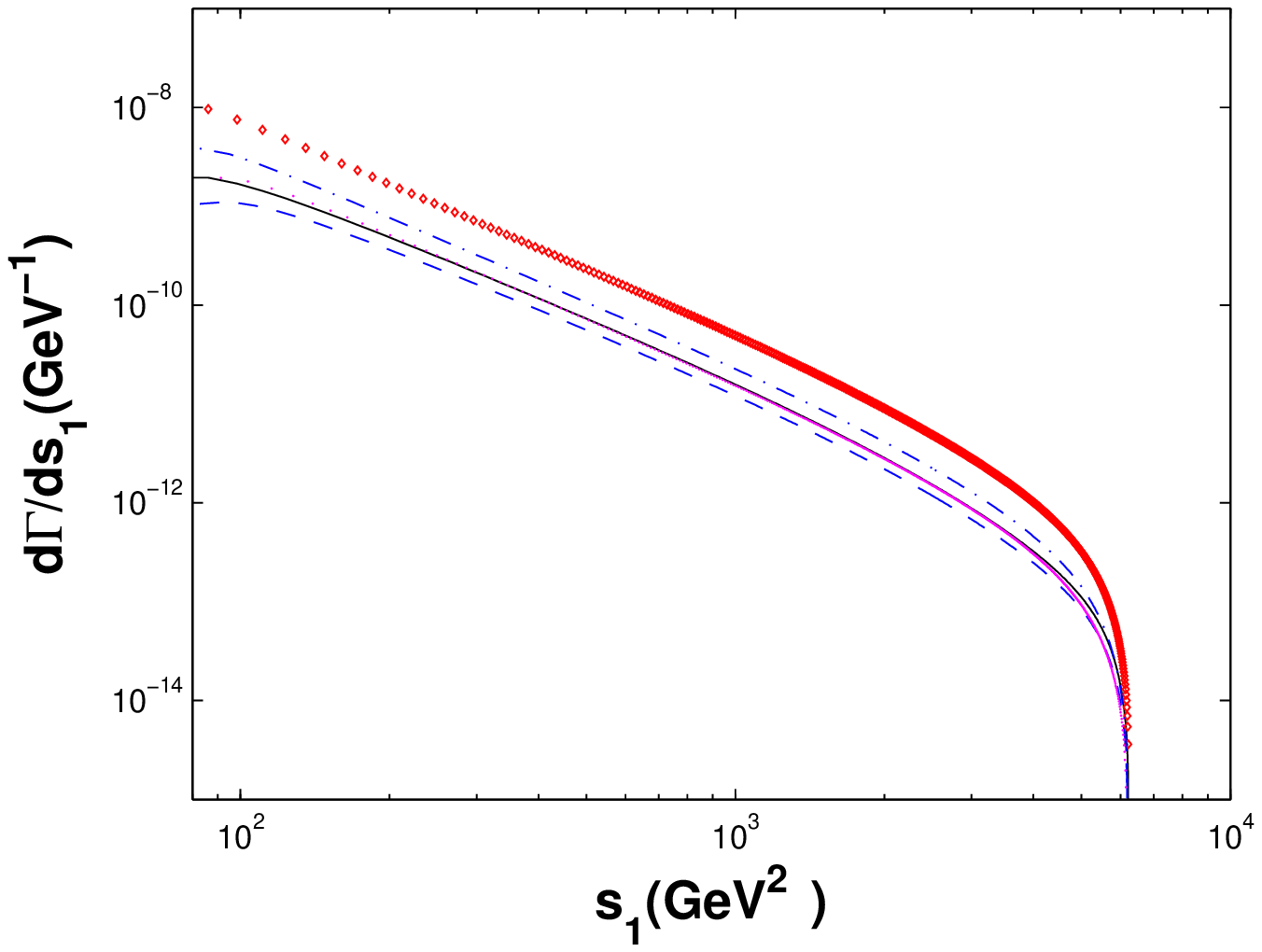}
\includegraphics[width=0.23\textwidth]{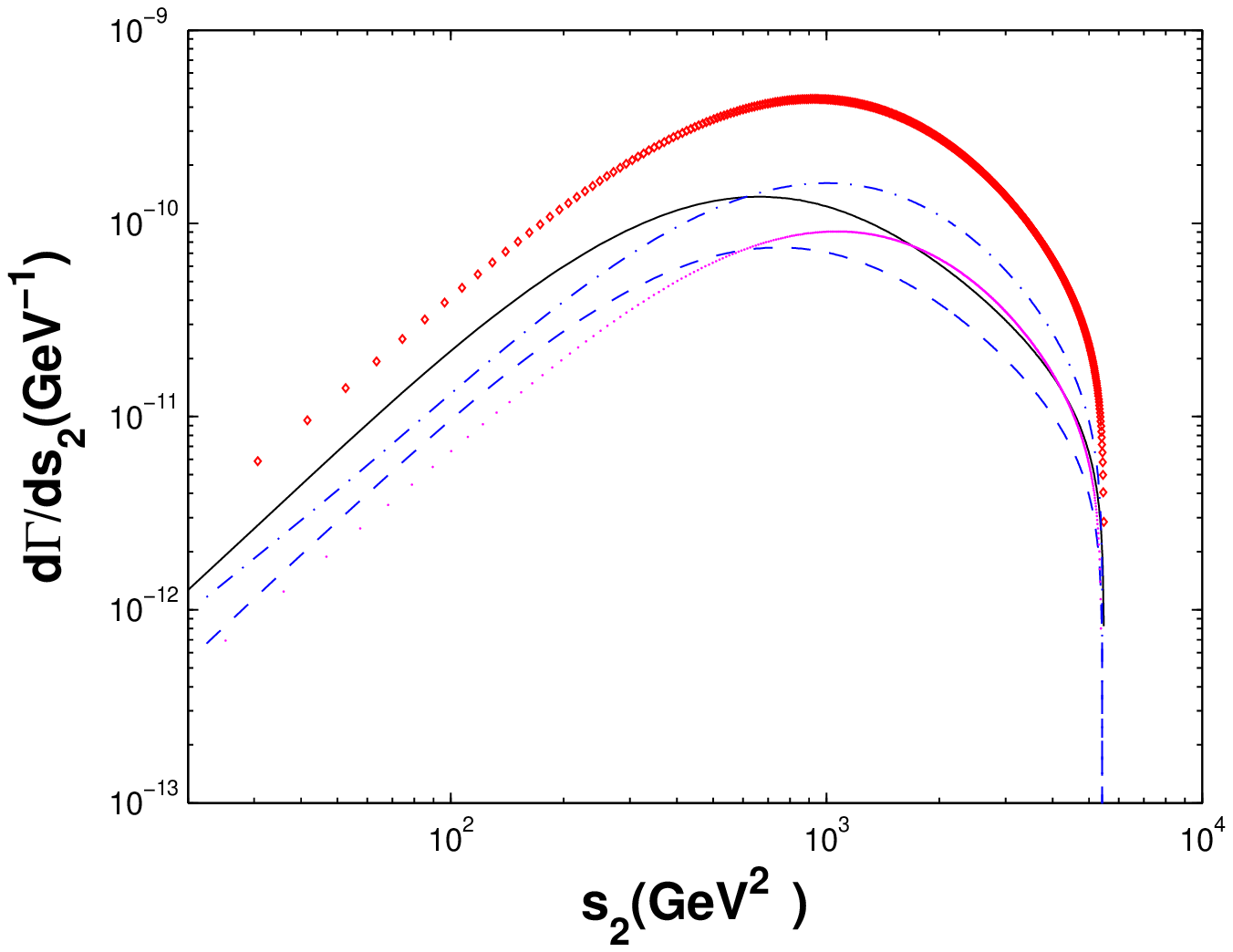}
\caption{Differential decay widths $d\Gamma/ds_1$ and $d\Gamma/ds_2$ for $W^+\rightarrow |(c\bar{b})[n]\rangle +c\bar{c}$, where the diamond, the dash-dot, the dotted, the solid and the dashed lines are for $|(c\bar{b})[1S]\rangle$, $|(c\bar{b})[2S]\rangle$, $|(c\bar{b})[3S]\rangle$, $|(c\bar{b})[1P]\rangle$ and $|(c\bar{b})[2P]\rangle$, respectively.} \label{Bcccdiss1s2}
\end{figure}

\begin{figure}
\includegraphics[width=0.23\textwidth]{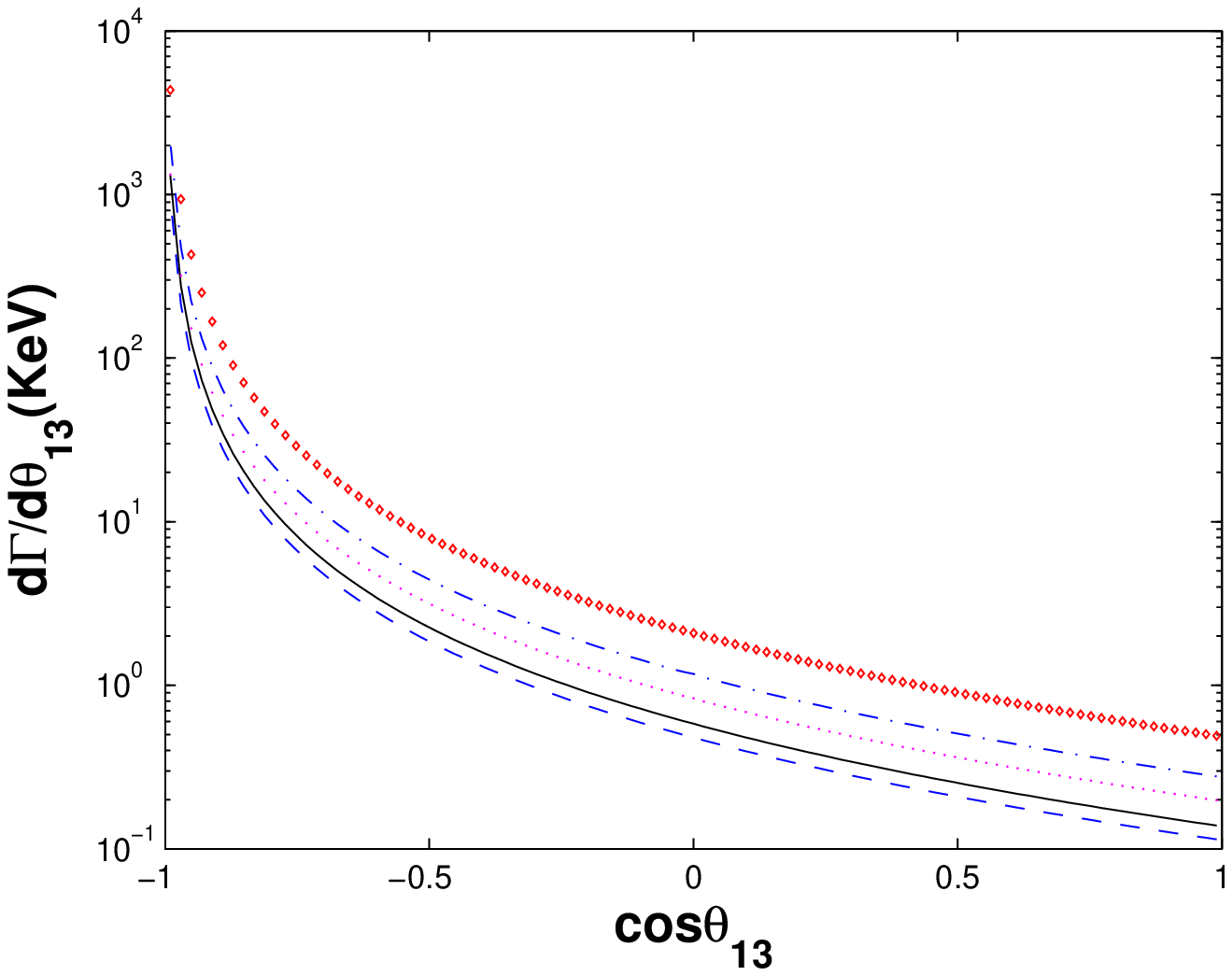}
\includegraphics[width=0.23\textwidth]{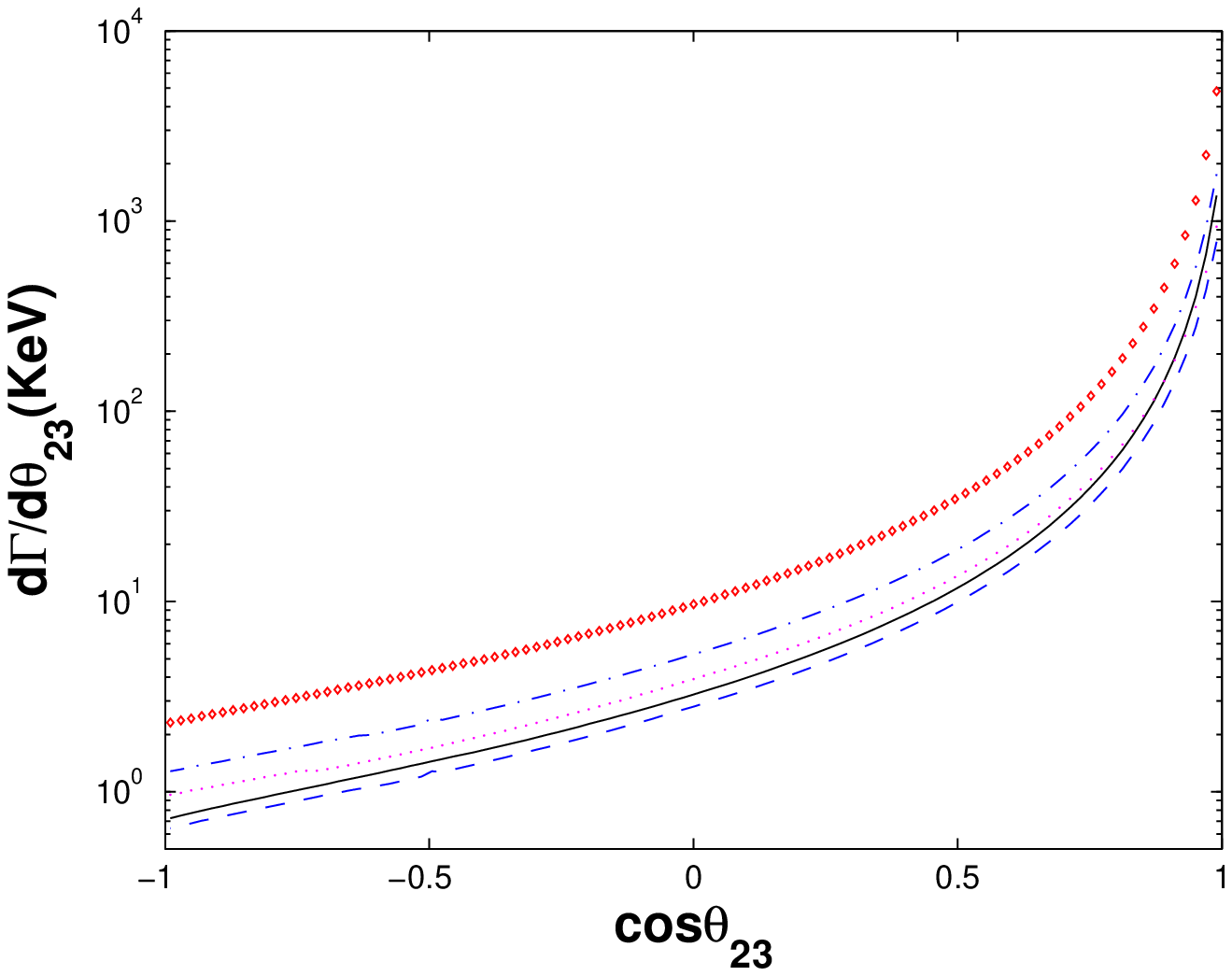}
\caption{Differential decay widths $d\Gamma/d\cos{\theta_{13}}$ and $d\Gamma/d\cos{\theta_{23}}$ for $W^+\rightarrow |(c\bar{c})[n]\rangle +c\bar{s}$, where the diamond, the dash-dot, the dotted, the solid and the dashed lines are for $|(c\bar{c})[1S]\rangle$, $|(c\bar{c})[2S]\rangle$, $|(c\bar{c})[3S]\rangle$, $|(c\bar{c})[1P]\rangle$ and $|(c\bar{c})[2P]\rangle$, respectively.} \label{CCcsdcos}
\end{figure}

\begin{figure}
\includegraphics[width=0.23\textwidth]{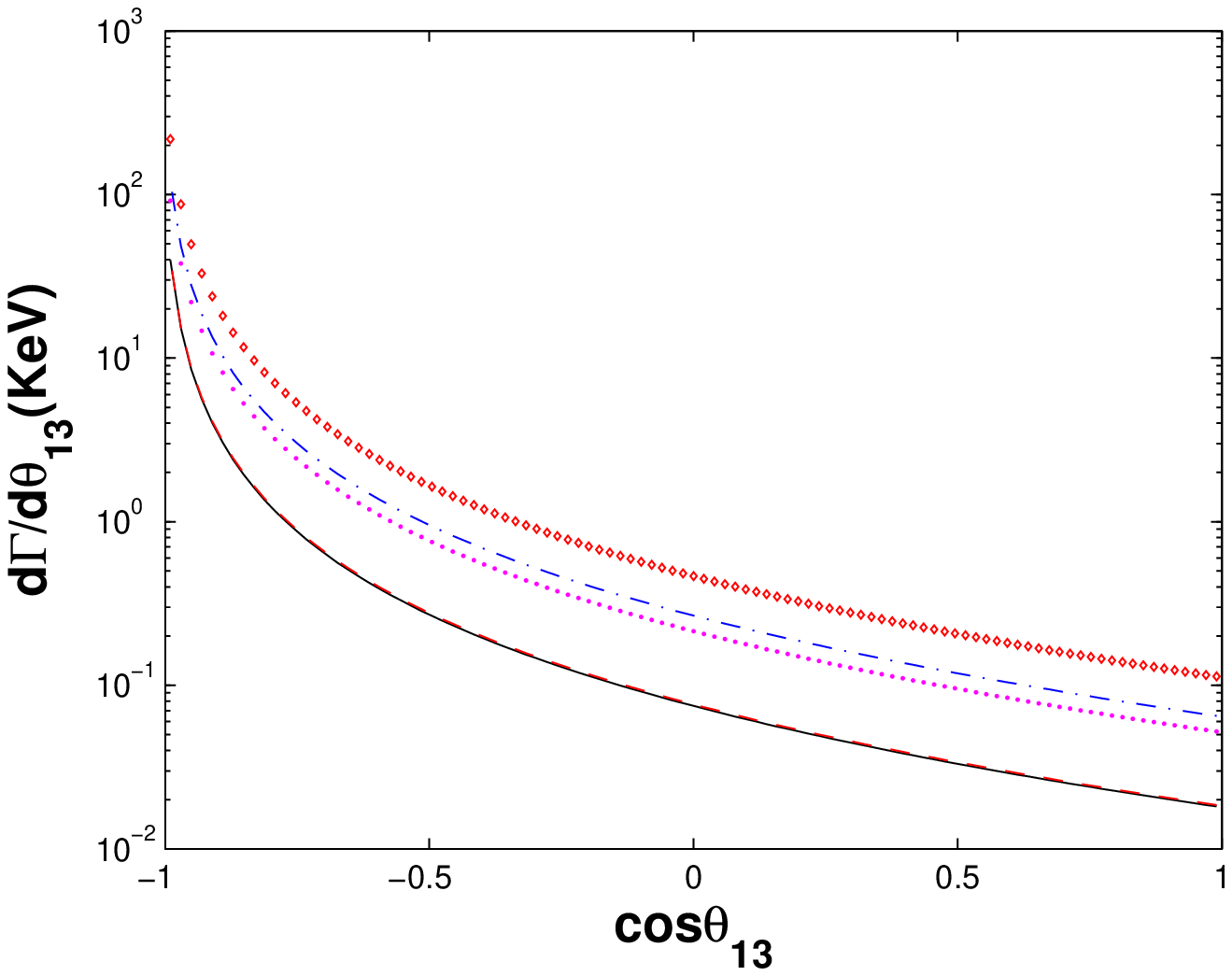}
\includegraphics[width=0.23\textwidth]{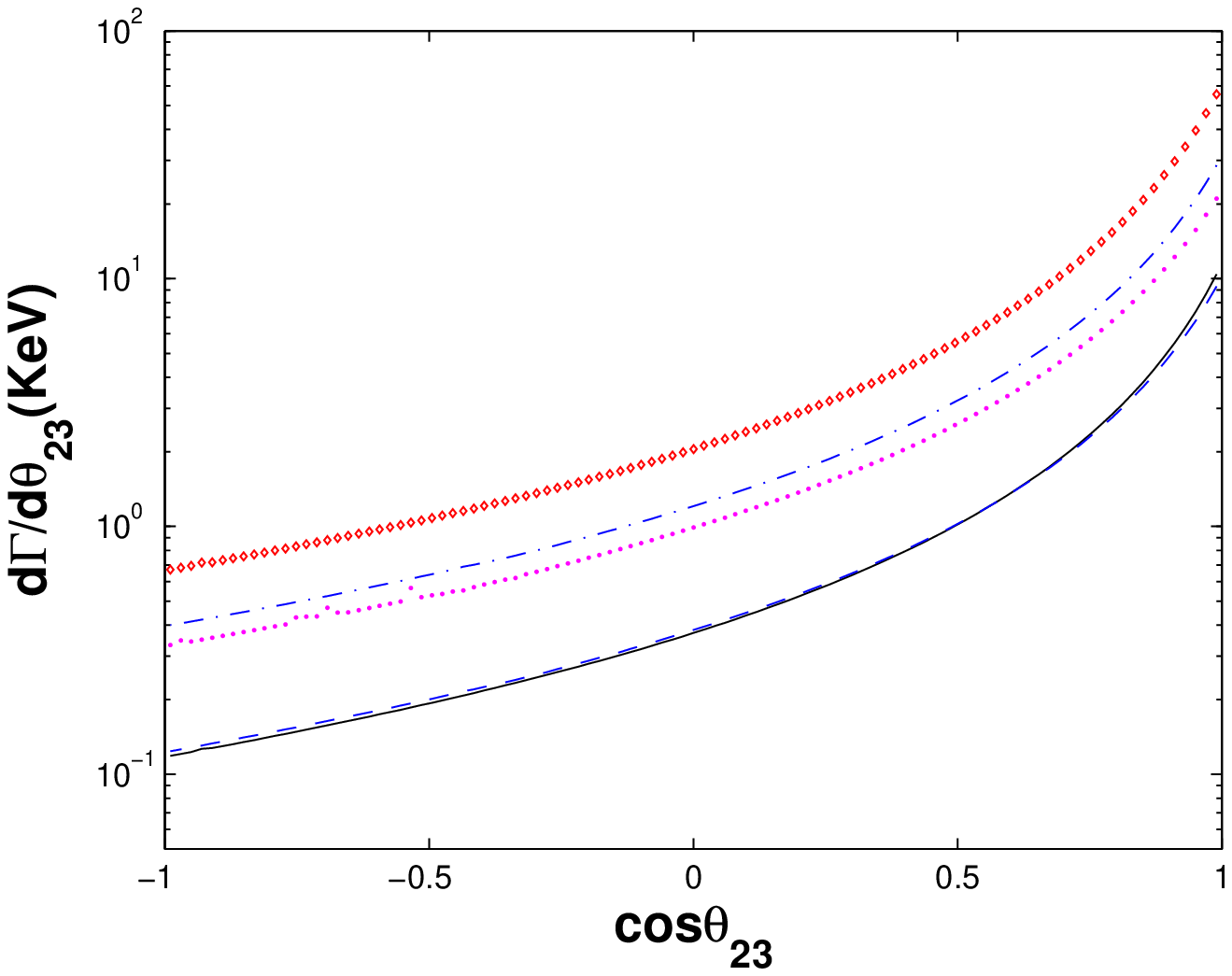}
\caption{Differential decay widths $d\Gamma/d\cos{\theta_{13}}$ and $d\Gamma/d\cos{\theta_{23}}$ for $W^+\rightarrow |(c\bar{b})[n]\rangle +b\bar{s}$, where the diamond, the dash-dot, the dotted, the solid and the dashed lines are for $|(c\bar{b})[1S]\rangle$, $|(c\bar{b})[2S]\rangle$, $|(c\bar{b})[3S]\rangle$, $|(c\bar{b})[1P]\rangle$ and $|(c\bar{b})[2P]\rangle$, respectively.} \label{Bcbsdcos}
\end{figure}

\begin{figure}
\includegraphics[width=0.23\textwidth]{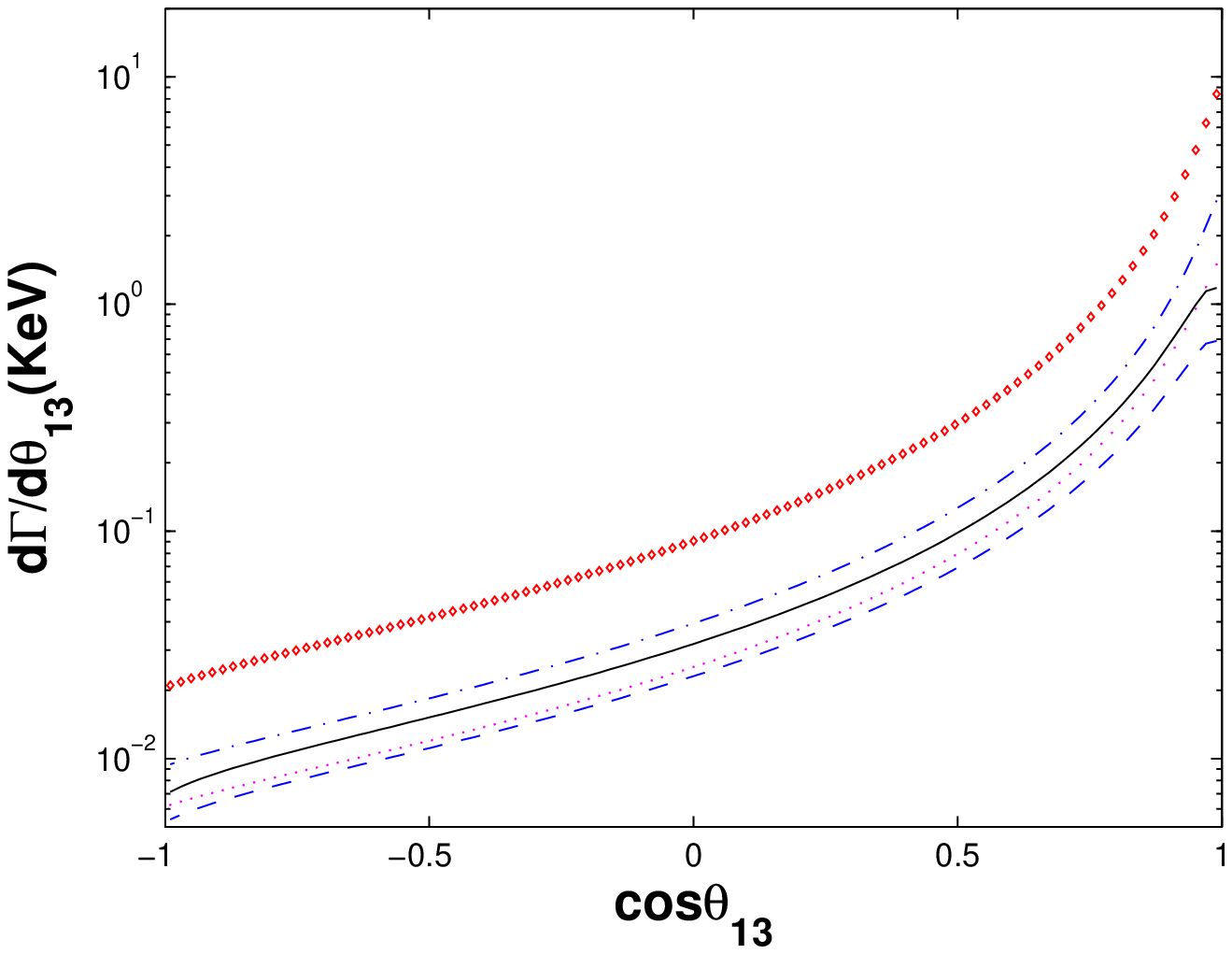}
\includegraphics[width=0.23\textwidth]{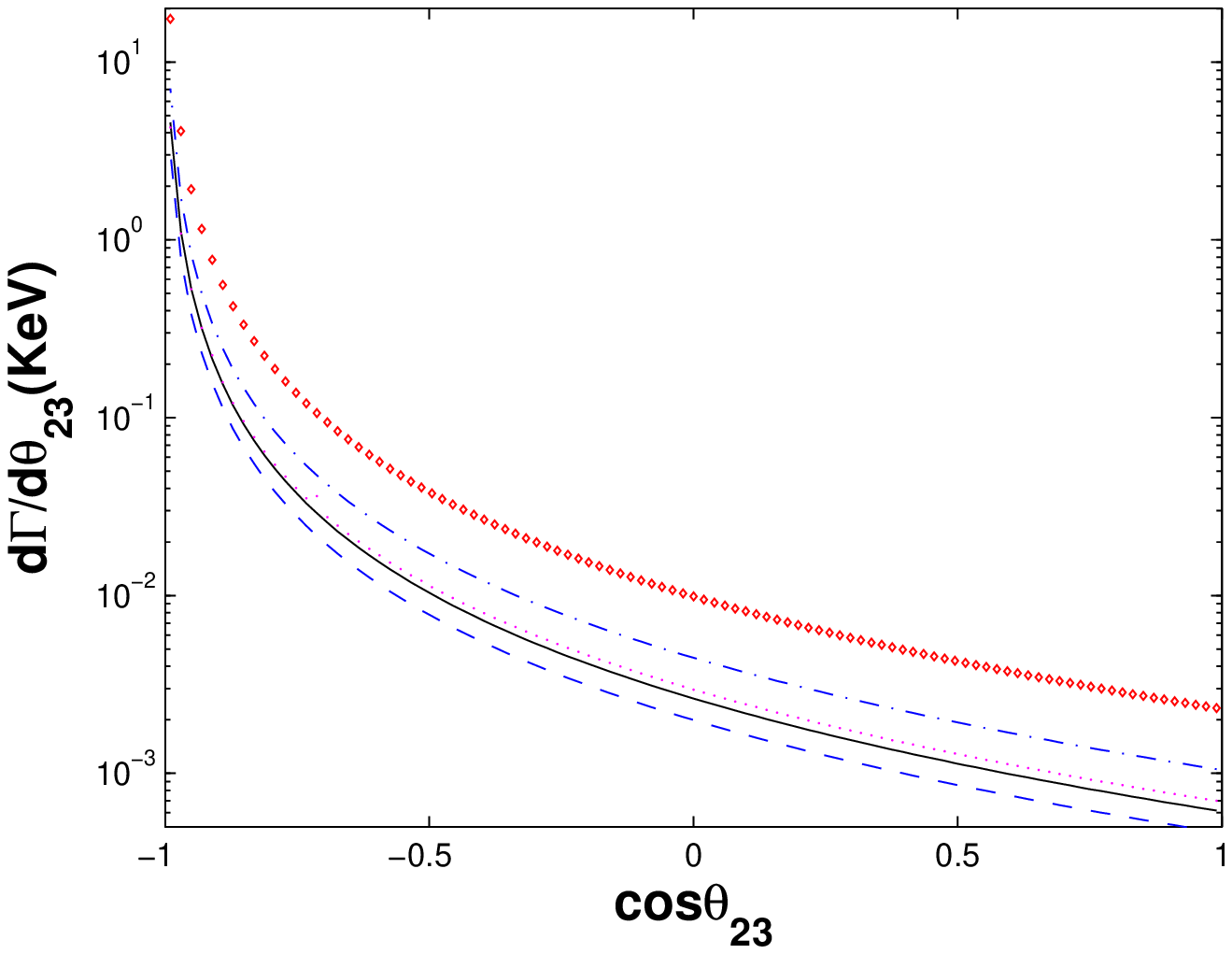}
\caption{Differential decay widths $d\Gamma/d\cos{\theta_{13}}$ and $d\Gamma/d\cos{\theta_{23}}$ for $W^+\rightarrow |(c\bar{b})[n]\rangle +c\bar{c}$, where the diamond, the dash-dot, the dotted, the solid and the dashed lines are for $|(c\bar{b})[1S]\rangle$, $|(c\bar{b})[2S]\rangle$, $|(c\bar{b})[3S]\rangle$, $|(c\bar{b})[1P]\rangle$ and $|(c\bar{b})[2P]\rangle$, respectively.} \label{Bccdcos}
\end{figure}

\begin{figure}[ht]
\includegraphics[width=0.23\textwidth]{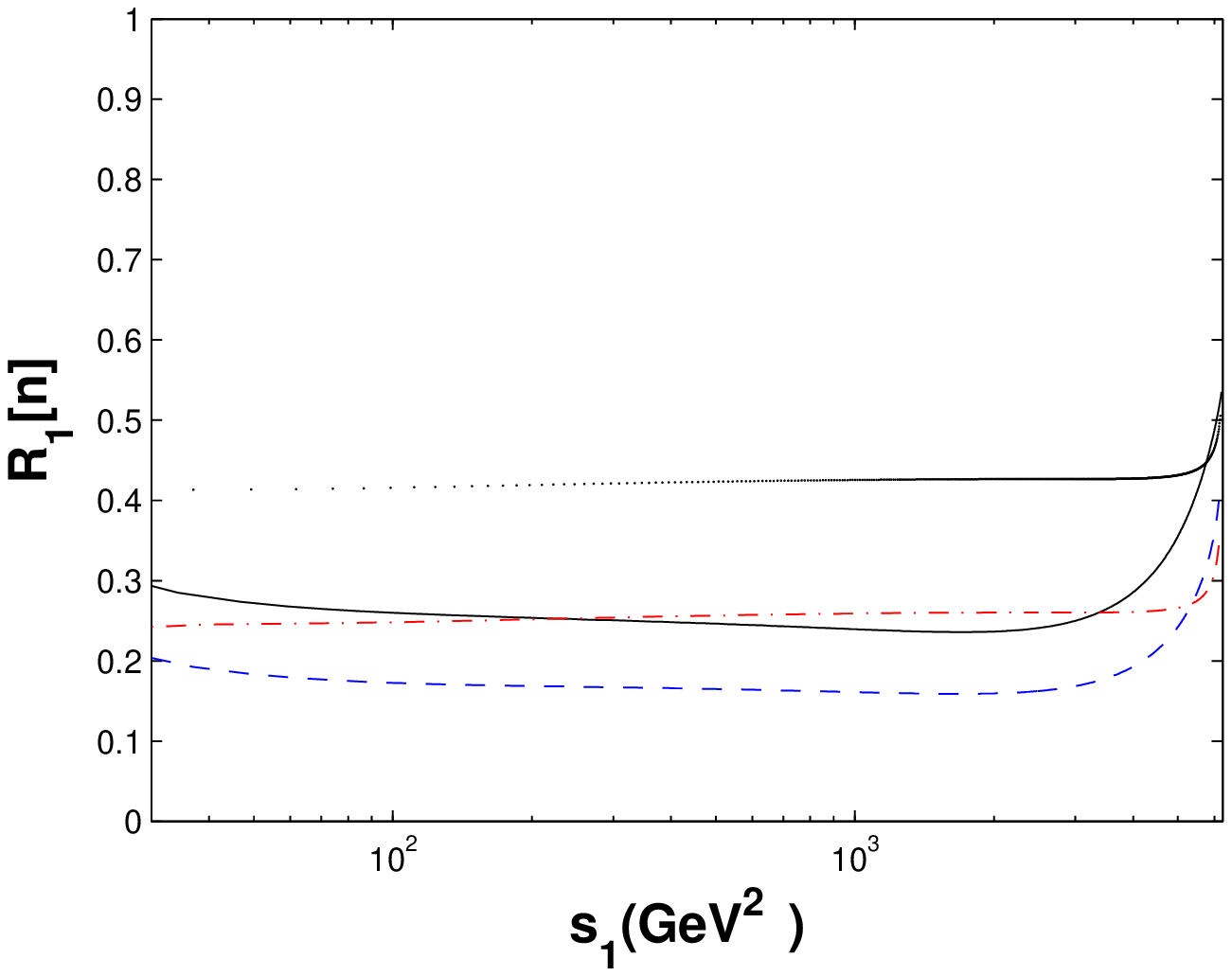}
\includegraphics[width=0.23\textwidth]{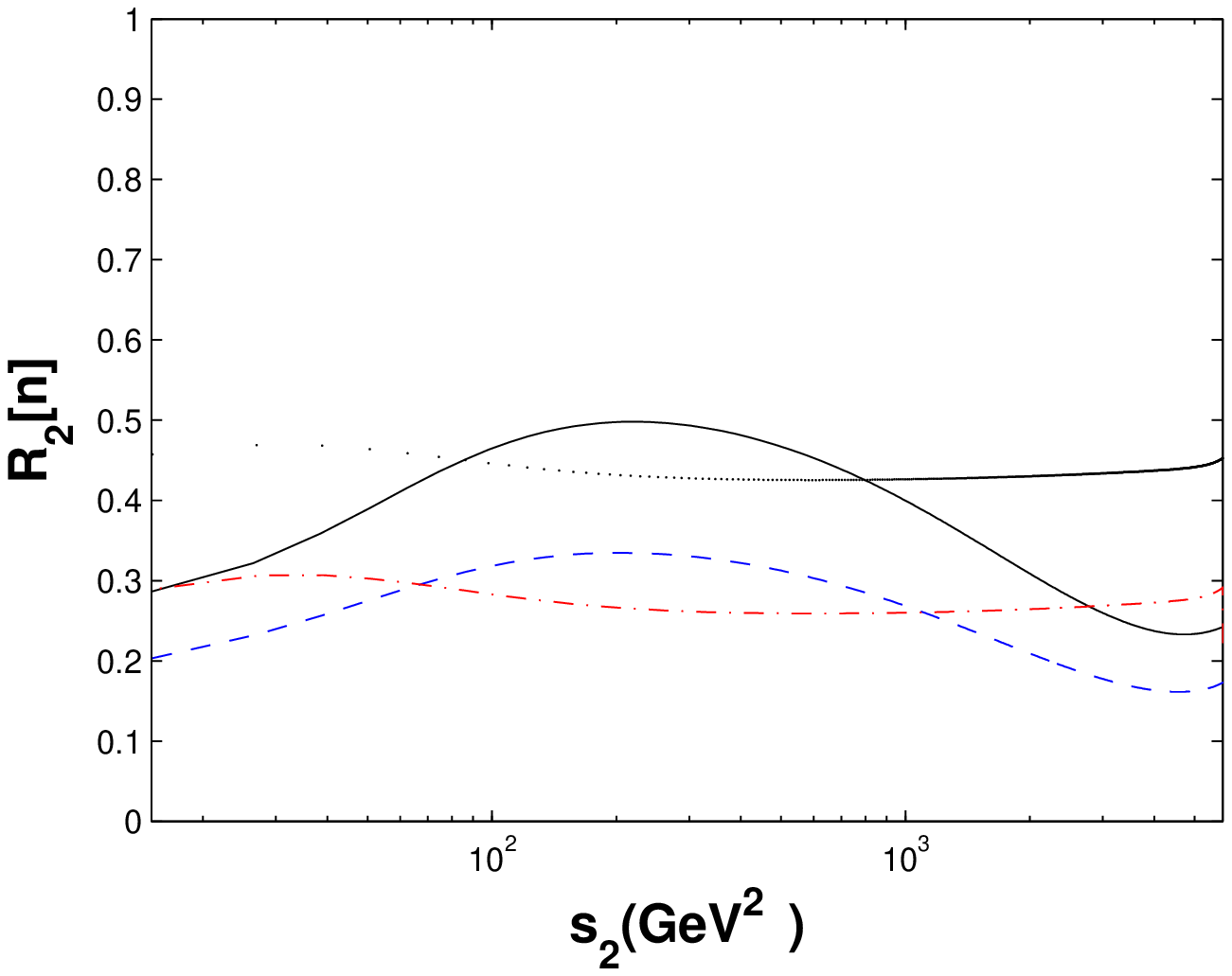}
\caption{The ratios $R_1[n]$ and $R_2[n]$ versus $s_1$ and $s_2$ for the channel $W^+\rightarrow |(c\bar{c})[n]\rangle +c\bar{s}$. Here the dotted, the dash-dot, the solid and the dashed lines are for $|(c\bar{c})[2S]\rangle$, $|(c\bar{c})[3S]\rangle$, $|(c\bar{c})[1P]\rangle$ and $|(c\bar{c})[2P]\rangle$, respectively. }  \label{CCcsdLds1ds2}
\end{figure}

To show the relative importance among different Fock-states more clearly, we present the differential distributions $d\Gamma/ds_1$, $d\Gamma/ds_2$, $d\Gamma/d\cos\theta_{13}$ and $d\Gamma/d\cos\theta_{12}$ for the mentioned channels in Figs.(\ref{CCcsdiss1s2},\ref{Bcbsdiss1s2},\ref{Bcccdiss1s2}) and Figs.(\ref{CCcsdcos},\ref{Bcbsdcos},\ref{Bccdcos}). Moreover, taking the channel $W^+\rightarrow |(c\bar{c})[n]\rangle +c\bar{s}$ as an example, we define a ratio
\begin{equation}
R_{i}[n]=\frac{d\Gamma/ds_{i} (|(c\bar{c})[n]\rangle)}{d\Gamma/ds_{i} (|(c\bar{c})[1S]\rangle)} ,
\end{equation}
where $i=1,2$. The curves are presented in Fig.(\ref{CCcsdLds1ds2}). These figures show explicitly that higher Fock-states $|(Q\bar{Q}')[2S]\rangle$, $|(Q\bar{Q}')[3S]\rangle$, $|(Q\bar{Q}')[1P]\rangle$ and $|(Q\bar{Q}')[2P]\rangle$ can provide sizable contributions in comparison to the lower Fock-state $|(Q\bar{Q}')[1S]\rangle$ in almost the whole kinematical region.

If assuming all the higher excited heavy-quarkonium states decay to the ground spin-singlet $S$-wave state ($(Q\bar{Q'})|[1{^1S_0}]\rangle$) with $100\%$ efficiency via electromagnetic or hadronic interactions, we obtain the total decay width
\begin{eqnarray}
\Gamma_{W^+\to |(c\bar{c})\rangle +c\bar{s}} &=& 591.3 \;{\rm KeV},\label{g1}\\
\Gamma_{W^+\to |(c\bar{b})\rangle +b\bar{s}} &=& 27.0 \;{\rm KeV},\label{g2}\\
\Gamma_{W^+\to |(c\bar{b})\rangle +c\bar{c}} &=& 2.01 \;{\rm KeV},\label{g3}\\
\Gamma_{W^+\to |(b\bar{b})\rangle +c\bar{b}} &=& 93.3\;{\rm eV} . \label{g4}
\end{eqnarray}
It shows that the total decay width of the bottomonium case is small as previously mentioned.

At the LHC, with the luminosity $\mathcal{L}_{p-p}=10^{34} \mathrm{cm}^{-2}\mathrm{s}^{-1}$ and the center-of-mass energy $\sqrt{s}=14$ TeV, about $3.07 \times10^{10}$ of $W^+$ events per year can be produced \cite{Jon}. Then, we can estimate the event numbers of $|(Q\bar{Q'})\rangle$-quarkonium production through $W^+$ decays; i.e. $3.95\times 10^6$ $|(c\bar{c})[1S]\rangle$, $1.69\times10^6$ $|(c\bar{c})[2S]\rangle$, $8.29\times10^5$ $|(c\bar{c})[3S]\rangle$, $1.39\times10^6$ $|(c\bar{c})[1P]\rangle$, $8.49\times10^5$ $|(c\bar{c})[2P]\rangle$ events per year can be produced; $1.90\times10^5$ $|(c\bar{b})[1S]\rangle$, $1.03\times10^5$ $|(c\bar{b})[2S]\rangle$, $7.91\times10^4$ $|(c\bar{b})[3S]\rangle$, $2.64\times10^4$ $|(c\bar{b})[1P]\rangle$, $2.91\times10^4$ $|(c\bar{b})[2P]\rangle$ events per year can be produced.

\subsection{Decay widths under four potential models}

\begin{table}
\begin{tabular}{|c||c|c|c|c|}
\hline
~~~~ & ~~B.T.\cite{wb}~~ & ~~P.L.\cite{am}~~ & ~~ Log.\cite{cj}~~ & ~~Cor.\cite{ektk} ~~ \\
\hline
$[n]= [1{^1S_{0}}]$  &132.0&162.8&132.8&236.9\\
$[n]= [1{^3S_{1}}]$  &136.4&168.3&137.3&244.9\\
\hline
$[n]= [2{^1S_{0}}]$ &56.36&59.55&44.53&98.76\\
$[n]= [2{^3S_{1}}]$ &58.30&61.60&46.06&102.2\\
\hline
$[n]= [3{^1S_{0}}]$ &22.85&20.59&14.36&39.72\\
$[n]= [3{^3S_{1}}]$ &33.38&30.08&20.98&58.04\\
\hline\hline
$[n] = [1{^1P_{1}}]$ &22.95&38.25&23.87&40.09\\
$[n] = [1{^3P_{0}}]$ &28.33&47.22&29.47&49.49\\
$[n] = [1{^3P_{1}}]$ &28.31&47.18&29.44&49.45 \\
$[n] = [1{^3P_{2}}]$ &14.76&24.59&15.35&25.77\\
\hline
$[n] = [2{^1P_{1}}]$ &12.98&16.67&9.673&23.67\\
$[n] = [2{^3P_{0}}]$ &17.21&22.10&12.82&31.38\\
$[n] = [2{^3P_{1}}]$ &19.07&24.50&14.21 &34.78\\
$[n] = [2{^3P_{2}}]$ &8.414&10.81&6.269&15.34\\
\hline\hline
~~sum~~ &591.3&734.2&537.1&1050.\\
\hline
\end{tabular}
\caption{Decay widths (in unit: KeV) for $(c\bar{c})$-charmonium production channel $ W^+\rightarrow |(c\bar{c})[n]\rangle +c \bar{s}$, where bound-state parameters from four potential models are adopted. }
\label{ccmm}
\end{table}

\begin{table}
\begin{tabular}{|c||c|c|c|c|}
\hline
~~~~ &~~B.T.\cite{wb}~~&~~P.L.\cite{am}~~&~~Log.\cite{cj}~~&~~Cor.\cite{ektk}~~ \\
\hline
$[n]= [1{^1S_{0}}]$ &6.39&6.66&5.87&12.4 \\
$[n]= [1{^3S_{1}}]$ &5.49&5.72&5.05&10.7  \\
\hline
$[n]= [2{^1S_{0}}]$ &3.53&3.41&2.76&6.33 \\
$[n]= [2{^3S_{1}}]$&3.07&2.97&2.40&5.51 \\
\hline
$[n]= [3{^1S_{0}}]$&2.74&2.28&1.89&4.84 \\
$[n]= [3{^3S_{1}}]$ &2.41&2.01&1.66&4.26\\
\hline\hline
$[n] = [1{^1P_{1}}]$ &0.270&0.439&0.321&0.459 \\
$[n] = [1{^3P_{0}}]$&0.733&1.19&0.871&1.25\\
$[n] = [1{^3P_{1}}]$&0.514&0.836&0.611&0.875\\
$[n] = [1{^3P_{2}}]$&0.0336&0.0547&0.0400&0.0572\\
\hline
$[n] = [2{^1P_{1}}]$&0.360&0.480&0.326&0.629\\
$[n] = [2{^3P_{0}}]$&0.693&0.924&0.627&1.21\\
$[n] = [2{^3P_{1}}]$&0.724&0.965&0.655&1.26\\
$[n] = [2{^3P_{2}}]$&0.0392&0.0522&0.0355&0.0684\\
\hline\hline
~~sum~~ &27.0&28.0&23.1&49.8\\
\hline
\end{tabular}
\caption{Decay widths (in unit: KeV) for $(c\bar{b})$-quarkonium production through $W^+\rightarrow |(c\bar{b})[n]\rangle +b \bar{s}$, where bound-state parameters from four potential models are adopted. }
\label{cbmm}
\end{table}

\begin{table}
\begin{tabular}{|c||c|c|c|c|}
\hline
~~~ &~~B.T.\cite{wb}~~&~~P.L.\cite{am}~~&~~Log.\cite{cj}~~&~~Cor.\cite{ektk}~~ \\
\hline
$[n]= [1{^1S_{0}}]$ &0.411&0.428&0.378&0.798\\
$[n]= [1{^3S_{1}}]$ &0.593&0.618&0.545&1.15\\
\hline
$[n]= [2{^1S_{0}}]$&0.160&0.155&0.126&0.288\\
$[n]= [2{^3S_{1}}]$ &0.224&0.216&0.175&0.402\\
\hline
$[n]= [3{^1S_{0}}]$ &0.0942&0.0784&0.0649&0.166\\
$[n]= [3{^3S_{1}}]$ &0.128&0.107&0.0883&0.227\\
\hline\hline
$[n] = [1{^1P_{1}}]$ &0.105&0.170&0.124&0.178\\
$[n] = [1{^3P_{0}}]$&0.0255&0.0415&0.0303&0.0434\\
$[n] = [1{^3P_{1}}]$&0.0536&0.0872&0.0637&0.0912\\
$[n] = [1{^3P_{2}}]$&0.0598&0.0972&0.0711&0.102\\
\hline
$[n] = [2{^1P_{1}}]$&0.0659&0.0878&0.0596&0.115\\
$[n] = [2{^3P_{0}}]$&0.0185&0.0247&0.0168&0.0324\\
$[n] = [2{^3P_{1}}]$&0.0364&0.0486&0.0330&0.0636\\
$[n] = [2{^3P_{2}}]$&0.0374&0.0498&0.0338&0.0653\\
\hline\hline
~~sum~~ &2.01&2.21&1.81&3.72\\
\hline
\end{tabular}
\caption{Decay widths (in unit: KeV) for $(c\bar{b})$-quarkonium production  through $W^+\rightarrow |(c\bar{b})[n]\rangle +c \bar{c}$, where bound-state parameters from four potential models are adopted. }
\label{cbmmm}
\end{table}

Next, we discuss the uncertainty caused by the bound-state parameters. These parameters are main uncertainty sources for estimating the heavy $|(Q\bar{Q'})\rangle$-quarkonium production. We take the parameters derived under four potential models, i.e. the Buchm\"{u}ller-Tye model \cite{wb}, the Power-Law model \cite{am}, the Logarithmic model \cite{cj} and the Cornell model \cite{ektk}, to do our discussion. We take the symbols B.T., P.L., Log. and Cor. as short notations for the Buchm\"{u}ller-Tye model, the Power-Law model, the Logarithmic model and the Cornell model, respectively. The constitute quark masses and their corresponding radial wavefunctions at the origin and the first derivative of the radial wavefunctions at the origin for the $|(Q\bar{Q'})\rangle$-quarkonium states can be found in Tables I, II and III of Ref.\cite{quigg}. To short the paper, we do not repeat them here.

The decay width for $|(Q\bar{Q'})\rangle$-quarkonium production under four potential models are presented in Tables \ref{ccmm}, \ref{cbmm} and \ref{cbmmm}. The Cornell model, which is a naive Coulomb-plus-linear potential, always gives the largest values among the four models. While, the decay widths for the other three models are consistent with each other: taking the BT-model decay width as the center value, for the channel $ W^+\rightarrow |(c\bar{c})[n]\rangle +c \bar{s}$, we obtain the uncertainty $\left(^{+24\%}_{-9\%}\right)$, where the upper value is for the Power-Law model and the lower value is for the Logarithmic model; for the channel $W^+\rightarrow |(c\bar{b})[n]\rangle +b \bar{s}$, we obtain the uncertainty $\left(^{+4\%}_{-15\%}\right)$; for the channel $W^+\rightarrow |(c\bar{b})[n]\rangle +c \bar{c}$, we obtain the uncertainty $\pm 10\%$.

\section{summary}

In addition to our previous studies on the lower Fock states' production presented in Ref.\cite{wsp}, we have carried out an investigation on the higher-excited $|(Q\bar{Q})\rangle$-quarkonium states' production through $W^+$-boson semi-inclusive decays.

For the $|(c\bar{c})\rangle$-charmonium production channel $W^{+}\rightarrow |(c\bar{c})[n]\rangle + c \bar{s}$, the decay width for $[n]=2S$, $3S$, $1P$ and $2P$ states are about $43\%$, $21\%$, $35\%$ and $21\%$ of that of the $1S$-level; For the $|(c\bar{b})\rangle$-quarkonium production channel $W^{+}\rightarrow |(c\bar{b})[n]\rangle + b\bar{s}$, the decay width for $[n]=2S$, $3S$, $1P$ and $2P$ states are about $55\%$, $43\%$, $13\%$ and $15\%$ of that of the $1S$-level; For the $|(c\bar{b})\rangle$-quarkonium production channel $W^{+}\rightarrow |(c\bar{b})[n]\rangle + c\bar{c}$, the decay width for $[n]=2S$, $3S$, $1P$ and $2P$ states are about $38\%$, $22\%$, $24\%$ and $16\%$ of that of the $1S$-level. Then these higher-excited $nS$-level and $nP$-level states with $n\geq 2$ can also provide sizable contributions to the heavy quarkonium production. Therefore, we need to take these higher excited states into consideration for a sound estimation.

At the LHC with the luminosity ${\cal L}\propto 10^{34}cm^{-2}s^{-1}$ and the center-of-mass energy $\sqrt{S}=14$ TeV, with the help of Eqs.(\ref{g1},\ref{g2},\ref{g3},\ref{g4}), one can obtain $8.7\times10^6$ $\eta_c$ and $J/\Psi$, $4.3\times10^5$ $B_c$ and $B^*_c$ and $1.4\times10^3$ $\eta_b$ and $\Upsilon$ events per year. Uncertainties caused by the bound-state parameters are presented, four potential models are adopted for the purpose. It is found that the decay widths for the BT model, the Power-Law model and the Logarithmic are consistent with each other, whereas the naive Cornell model is not.

\hspace{2cm}

{\bf Acknowledgements}: This work was supported in part by the Fundamental Research Funds for the Central Universities under Grant No.CDJXS1102209, the Program for New Century Excellent Talents in University under Grant No.NCET-10-0882, and the Natural Science Foundation of China under Grant No.11075225.


\begin{thebibliography}{99}

\bibitem{nrqcd} G.T. Bodwin, E. Braaten and G.P. Lepage, Phys. Rev. D {\bf 51}, 1125 (1995); Erratum Phys. Rev. D {\bf 55}, 5853 (1997).

\bibitem{wsp}  Q.L. Liao, X.G. Wu, J. Jiang, Z. Yang and Z.Y. Fang, Phys.Rev. D{\bf 85}, 014032 (2012).

\bibitem{w} C.F. Qiao, L.P. Sun, D.S. Yang and R.L. Zhu, Eur.Phys.J. C{\bf 71}, 1766 (2011).

\bibitem{projector1} R. Baier and R. R\"{u}ckl, Z.Phys. C{\bf19} 251 (1983); B. Humpert, Phys.Lett. B{\bf 184}, 105 (1987); R. Gastmans, W. Troost and T.T. Wu, Nucl.Phys. B {\bf 291}, 731 (1987); Y.Q. Chen, Phys.Rev. D{\bf 48}, 5181 (1993).

\bibitem{projector2} N. Brambilla {\it et al.}, Quarkonium Working Group Collaboration, hep-ph/0412158; N. Brambilla {\it et al.}, Eur.Phys.J. C{\bf 71}, 1534 (2011); N. Brambilla, A. Pineda, J. Soto and A. Vairo, Rev.Mod.Phys. {\bf 77}, 1423(2005).

\bibitem{projector3} A. Petrelli, M. Cacciari, M. Greco, F. Maltoni and M.L. Mangano, Nucl.Phys. B{\bf 514}, 245(1998).

\bibitem{itt0} R. Kleiss and W.J. Stirling, Nucl.Phys. B{\bf 262}, 235 (1985).

\bibitem{itt1} C.H. Chang and Y.Q. Chen, Phys.Rev. D{\bf 46}, 3845 (1992).

\bibitem{itt2} C.H. Chang, J.X. Wang and X.G. Wu, Phys.Rev. D{\bf 77}, 014022 (2008); X.G. Wu, Phys.Lett. B{\bf 671}, 318 (2009).

\bibitem{itt3} L.C. Deng, X.G. Wu, Z. Yang, Z.Y. Fang and Q.L. Liao, Eur.Phys.J. C{\bf 70}, 113(2010); Z. Yang, X.G. Wu, L.C. Deng, J.W. Zhang and G. Chen, Eur.Phys.J. C{\bf 71}, 1563(2011).

\bibitem{itt4} Z. Yang, X.G. Wu, G. Chen, Q.L. Liao and J.W. Zhang, Phys.Rev. D{\bf 85}, 094015 (2012).

\bibitem{petrelli} A. Petrelli, M. Cacciari, M. Greco, F. Maltoni and M.L. Mangano, Nucl.Phys. B{\bf 514}, 245(1998).

\bibitem{vegas} G.P. Lepage, J. Comp. Phys {\bf 27}, 192 (1978).

\bibitem{bcvegpy} C.H. Chang, C. Driouich, P. Eerola and X.G. Wu, Comput. Phys. Commun. {\bf 159}, 192 (2004); C.H. Chang, J.X. Wang and X.G. Wu, Comput.Phys.Commun. {\bf 174}, 241 (2006); C.H. Chang, J.X. Wang and X.G. Wu, Comput. Phys. Commun. {\bf 175}, 624 (2006); X.Y. Wang and X.G. Wu, Comput. Phys. Commun. {\bf 183}, 442 (2012).

\bibitem{genxicc} C.H. Chang, J.X. Wang and X.G. Wu, Comput. Phys. Commun.{\bf 177}, 467 (2007); Comput. Phys. Commun.{\bf 181}, 1144 (2010).

\bibitem{geg} G.T. Bodwin, E. Braaten and G.P. Lepage, Phys.Rev. D{\bf 51}, 1125 (1995); {\bf 55}, 5853 (E) (1997).

\bibitem{pdg} K. Nakamura,{\it et al.}, Particle Data Group, J.Phys. G{\bf 37}, 075021(2010).

\bibitem{wtd} J. Alcaraz,{\it etal.}, ariXiv:0911.2604.

\bibitem{pmc} S.J. Brodsky and L.D. Giustino, arXiv: 1107.0338; S.J. Brodsky and X.G. Wu, Phys.Rev. D{\bf 85}, 034038(2012); S.J. Brodsky and X.G. Wu, arXiv:1204.1405; S.J. Brodsky and X.G. Wu, arXiv:1203.5312.

\bibitem{quigg} E.J. Eichten and C. Quigg, Phys.Rev. D{\bf 52}, 1726 (1995).

\bibitem{wb} W. Buchm\"{u}ller and S.H.H. Tye, Phys.Rev. D{\bf 24}, 132 (1981).

\bibitem{am} A. Martin, Phys.Lett. B{\bf 93}, 338(1980).

\bibitem{cj} C. Quigg and J.L. Rosner, Phys.Lett. B{\bf 71}, 153(1977).

\bibitem{ektk} E. Eichten, K. Gottfried, T. Kinoshita, K.D. Lane, and T.M. Yan, Phys.Rev. D{\bf17}, 3090 (1978); 21, 313(E) (1980); 21, 203 (1980).

\bibitem{Jon} J.R. Gaunt, C.H. Kom, A. Kulesza and W. James Stirling, Eur. Phys.J. C{\bf 69}, 53(2010).

\end{thebibliography}
\end{document}